\documentclass[a4paper,11pt]{article}
\pdfoutput=1 

\usepackage{jheppub} 

\usepackage[T1]{fontenc} 

\usepackage{scalerel}
\usepackage{color}
\usepackage{latexsym}
\usepackage{mathrsfs}
\usepackage{amssymb}
\usepackage{amsmath}
\usepackage{amsfonts, epsf,epsfig,color}
\usepackage{graphicx}
\usepackage{tensor}
\usepackage{manfnt}
\usepackage{slashed}
\usepackage[all]{xy}
\include{macro}

\newcommand{\C}{\mathbb{C}}

\newcommand{\R}{\mathbb{R}}
\renewcommand{\P}{\mathbb{P}}

\newcommand{\T}{\mathbb{T}}

\newcommand{\e}{\mathrm{e}}

\renewcommand{\P}{\mathbb{P}}

\newcommand{\rd}{\, \mathrm{d}}

\newcommand{\be}{\begin{equation}\label}
\newcommand{\ee}{\end{equation}}
\newcommand{\bea}{\begin{eqnarray}\label}
\newcommand{\eea}{\end{eqnarray}}

\title{\textbf{An Alternative Perspective on \\Ambitwistor String Theory}}

\author{N. Carabine$^*$ and R. A. Reid-Edwards$^{\dagger}$}

\affiliation{$^*$E. A. Milne Centre for Astrophysics, \\University of Hull, HU6 7RX, United Kingdom}
\affiliation{}
\affiliation{$^\dagger$Department of Applied Mathematics and Theoretical Physics, \\University of Cambridge, CB3 0WA, United Kingdom}

\emailAdd{$^{\dagger}$r.a.reid-edwards@damtp.cam.ac.uk}

\abstract{We review some ideas on the relationship between massless superparticles and the division algebras to provide a new perspective on ambitwsitor string theories. The key concern is the critical theory. We show that this theory has a reducible soft algebra, rather than a conventional Lie algebra. This algebra only closes on-shell. The BV procedure is employed to deal with the on-shell closure of the  algebra and the classical Master Action is presented.
}

\begin{document} 
\maketitle

\flushbottom

\section{Introduction}\label{introduction}

Recent years have witnessed considerable success in applying ideas from twistor theory to the calculation of observables in four-dimensional massless supersymmetric field theory\footnote{See \cite{Witten:2003nn,Adamo:2011pv,ArkaniHamed:2012nw,Atiyah:2017erd} and references therein for a selective overview.}. In many ways twistors are the ideal variables to think about certain problems in supersymmetric field theory and it is natural to try to extend these methods to higher dimensions where new applications may be found. There is something special about four dimensional twistor theory and there are a number of ways to generalise Penrose's notion of a twistor \cite{Penrose:1967wn} to higher dimensions but none that enjoys all of the features explicit in four dimensions. Nonetheless any formalism that retains even some of the magic of four-dimensional twistor theory is likely to be worth pursuing.

In four dimensions solutions to the massless equations of motion of helicity $h$ are given by $(0,1)$-forms; elements of\footnote{Or equivalently using {\v C}ech cohomology.} $H_{\bar{\partial}}^{0,1}({\cal O}(-2h-2);\P\T)$, where twistor space $\P\T$ in this context is an open subset of $\C\P^3$. A natural choice and one that is in many ways closest to the original spirit of the twistor programme is to define twistor space for $d$ dimensions as the space of projective pure spinors of the complexified conformal group $SO(d+2;\C)$. Of particular interest beyond four dimensions is the case of $d=6$, where twistor space is a quadric\footnote{The condition $Z^2=0$ comes from the purity requirement.} inside $\C\P^7$ and progress has been made in studying free self-dual conformal theories there \cite{Mason:2011nw,Mason:2012va,Saemann:2011nb}. Physical states are given by\footnote{Corresponding to direct and indirect Penrose transforms respectively.} $H^2$ and $H^3$ cohomology classes (i.e. Dolbeault $(0,2)$ or $(0,3)$ -forms modulo exact ones) and it is not clear how to introduce interactions or how to describe non self-dual theories in such a framework. Matters only get worse in ten dimensions where elements of $H^5$ and $H^{10}$ cohomology classes describe physical states and the purity condition becomes even more cumbersome to deal with. Moreover, the natural connection with the space of null geodesics is lost.

Alternatively one can seek to generalise the study of the space of null geodesics to higher dimensions. In this context, the space is usually referred to as ambitwistor space \cite{LeBrun,Isenberg:1978kk}. It has been noted on a number of occasions \cite{Evans:1987tm, Berkovits:1990yc, Cederwall:1992bi} that the division algebras provide an interesting unified guide for how to think about ambitwistor space in dimensions three, four, six and ten.  Based on the generalisation of $SL(2;\C)$ to $SL(2;\mathbb{K}_{d-2})$, where $\mathbb{K}_{1}=\R$, $\mathbb{K}_{2}=\C$, $\mathbb{K}_{4}=\mathbb{H}$, and $\mathbb{K}_{8}=\mathbb{O}$. This is compelling due to the existence of the isomorphisms $SL(2;\mathbb{K}_{d-2})\simeq SO(d-1,1)$, the Lorentz group in $d$ dimensions.

In four dimensions, we have $SL(2;\C)$, where a basis is given by $\sigma^{\mu}$, where $\sigma^0=1$ and $\sigma^i$ are the Pauli matrices. A null momentum is written as $P^{\mu}= \lambda^a\sigma^{\mu}_{a\dot{a}}\widetilde{\lambda}^{\dot{a}}$. This may be written in terms of the pair of complex-valued spinors $\lambda_a$ and $\widetilde{\lambda}_{\dot{a}}$, as $P_{a\dot{a}}=\lambda_a\widetilde{\lambda}_{\dot{a}}$.  Under $SO(2)$ transformations $(\lambda_a,\widetilde{\lambda}_{\dot{a}})\rightarrow (\lambda_a e^{i\theta},e^{-i\theta}\widetilde{\lambda}_{\dot{a}})$, the momentum is preserved. Introducing twistors $Z^I=(\omega_{\dot{\alpha}},\lambda^{\alpha})\in\C\P^3$ and dual twistors $W_I=(\widetilde{\lambda}^{\dot{\alpha}},\widetilde{\omega}_{\alpha})\in\widetilde{\C\P}^3$, where $\omega_{\dot{\alpha}}$ and $\widetilde{\omega}_{\alpha}$ are given by the incidence relations $\omega_{\dot{\alpha}}=X_{\dot{\alpha}\alpha}\lambda^{\alpha}$ and $\widetilde{\omega}_{\alpha}=X_{\dot{\alpha}\alpha}\widetilde{\lambda}^{\dot{\alpha}}$, the generator of these $SO(2)$ transformations may be written as
$$
U=\lambda^{\alpha}\tilde{\omega}_{\alpha}-\omega_{\dot{\alpha}}\widetilde{\lambda}^{\dot{\alpha}}\equiv Z\cdot W,
$$
where we take $\tilde{\omega}_{\alpha}$ and $\omega_{\dot{\alpha}}$ to be conjugate to $\lambda^{\alpha}$ and $\widetilde{\lambda}^{\dot{\alpha}}$ respectively. The condition $U=0$ is simply the standard statement that ambitwistor space is the quadric $Z\cdot W=0$ inside  $\C\P^3\times\widetilde{\C\P}^3$. A natural action for an ambitwistor string embedding into projective ambitwistor space $\P\mathbb{A}$ was given in \cite{Geyer:2014fka}
$$
S=\int_\Sigma W\hspace{-.06cm}\cdot\hspace{-.06cm} \bar{\partial}Z-W\hspace{-.06cm}\cdot\hspace{-.06cm} \bar{\partial}Z+e\, U,
$$
where $e$ is a Lagrange multiplier imposing $U=W\cdot Z=0$. The basic form of this construction generalises to dimensions six and ten, where there there is a corresponding division algebra isomorphism. 

In six dimensions the isomorphism $SL(2;\mathbb{H})\simeq SO(5,1)$ suggests a natural link to the Quaternions. As in four dimensions, Ambitwistor space is defined by a constraint surface inside some larger space.  The $SO(2)$ symmetry in four dimensions becomes an $SU(2)$ symmetry in six dimensions. These symmetry groups are the isometry groups of the group manifolds $S^1$ and $S^3$ and, extending to ten dimensions, a connection with $S^7$ was found, this being a somewhat special case as it is parallelizable yet not a Lie group (it is the sole compact example). The relationships to the Hopf fibrations $S^1\xrightarrow{S^0} S^1$,  $S^3\xrightarrow{S^1} S^2$, $S^7\xrightarrow{S^3} S^4$, and $S^{15}\xrightarrow{S^7} S^8$ (the projective spaces $\mathbb{R}\P^1$, $\mathbb{C}\P^1$, $\mathbb{H}\P^1$ and $\mathbb{O}\P^1$ respectively) were explored in \cite{Cederwall:1993nx}.

Our interests here are in the ten-dimensional case. Here the isomorphism $SL(2;\mathbb{O})\simeq SO(9,1)$ suggests a natural link to the Octonions. $S^7$ is a paralellizable but is not the manifold of a Lie group and, as noted in \cite{Cederwall:1992bi,Berkovits:1990yc}, the algebra of the ten-dimensional constraints is not a Lie algebra but is a \emph{soft} algebra in which the structure `constants' depend explicitly on $\lambda^a(z)$ fields and so are really structure \emph{functions}. This fact will add a complexity not present in the four- and six-dimensional cases. A key motivation for studying the ten-dimensional case is the recent work on understanding the origin of the CHY formulation of scattering amplitudes \cite{Cachazo:2013iea,Cachazo:2013hca} given by the ambitwistor string theory of \cite{Mason:2013sva} (see also \cite{Berkovits:2013xba}). The discussion here provides a unified description that includes the four-dimensional ambitwistor string of \cite{Geyer:2014fka} and the ten-dimensional one of \cite{Mason:2013sva}. It also may provide insight into why other ambitwistor strings have been less successful as critical string theories.

In the following section we discuss the geometry of the ambitwistor space and the construction of sigma models. Of particular interest are the constraints that must be imposed. In section three we construct the BRST charge; however, the failure of the gauge symmetry to close off-shell means that the naive BRST charge is not nilpotent when acting on the full space of fields. In section four we discuss gauge-fixing and employ the machinery of the BV formalism to enlarge the space of fields to allow for a nilpotent BRST transformation. The classical Master Action is presented and is found to require quadratic terms in the antifields. Section five briefly discusses outstanding issues and directions for future work.

\section{A Spinorial Perspective on Ambitwistor String Theory}

We start with the ambitwistor string of \cite{Mason:2013sva} with action
\begin{equation}\label{XP}
S=\int_{\Sigma}P_{\mu}\bar{\partial}X^{\mu}+\mu T+\frac{h}{2}P^2.
\end{equation}
The target space is the space of null geodesics. The stress tensor is $T(z)=P_{\mu}\partial X^{\mu}$ and the Beltrami differentials $\mu(z)$ and $h(z)$ are Lagrange multipliers which impose the conditions $T(z)=0$ and $P^2(z)=0$ respectively. One may think of this action as a holomorphic version of the massless particle worldline action and generalisations with manifest worldsheet or spacetime supersymmetry also exist \cite{Mason:2013sva,Berkovits:2013xba}. There is a gauge symmetry corresponding to the constraints and we gauge fix  $\mu(z)$ and $h(z)$ in the usual way and introduce holomorphic ghost systems, $(b,c)$ and $(\tilde{b},\tilde{c})$, both of conformal weight $2$. The theory is conformal in 26 (complex) dimensions but the physical interpretation of this theory is unclear\footnote{Although significant progress in our understanding has been made recently by \cite{Berkovits:2018jvm}.}; however, the supersymmetric generalisations appear to reproduce perturbative type II supergravity in ten-dimensions.

It is worth taking the time to stress that, unlike the superparticle, the ambitwistor string theory is a CFT and enjoys all of the privileges of the state-operator correspondence. As such, the physical states of the theory describe operator deformations of the target space. And unlike the worldline theory, which describes a particle moving on a potentially curved yet static background, the ambitwistor string is expected to encode the full dynamics of that background. Thus, despite superficial similarities in the formalism, these ambitwistor string theories are qualitatively distinct from their worldline counterparts.

Taking the action (\ref{XP}) as a starting point, we wish to recast it in a more manifestly spinorial form along the lines of that achieved in \cite{Berkovits:1990yc,Cederwall:1992bi} for the superparticle. On the constraint surface $P^2(z)=0$, the field $P_{\mu}(z)$ may be written as
\begin{equation}\label{P}
P^{\mu}(z)=\lambda^a(z)\,\Gamma^{\mu}_{ab}\,\lambda^b(z).
\end{equation}
where $a=1,2,3...(2d-4)$ and it is important to note the weight $1/2$ field $\lambda^a(z)$ is a generic spinor. It is \emph{not} pure. The $\Gamma_{ab}^{\mu}=\Gamma_{ba}^{\mu}$ satisfy the Clifford algebra relation $\{\Gamma^{\mu},\Gamma^{\nu}\}=2\eta^{\mu\nu}$ and are related to a representation of the ten-dimensional gamma matrices $\gamma_{\mu A}{}^{B'}$ by
$$
\gamma_{\mu A}{}^{B'}=\left(\begin{array}{cc}
0 & \Gamma^{ab}_{\mu}\\
\Gamma_{ab}^{\mu} & 0
\end{array}\right).
$$
The identity
\begin{equation}\label{identity}
\Gamma_{ab}^{\mu}\Gamma_{\mu cd}+\Gamma_{ad}^{\mu}\Gamma_{\mu bc}+\Gamma_{ac}^{\mu}\Gamma_{\mu db}=0,
\end{equation}
which holds for $d=3,4,6,10$, ensures that $\Gamma_{ab}^{\mu}P_{\mu}\lambda^b=0$. Hence the weaker condition $P^2(z)=0$ follows identically. The identity (\ref{identity}) is the crucial fact that makes the connection with the division algebra possible \cite{Evans:1987tm}. We introduce weight $1/2$ fields $\omega_a(z)$ via the incidence relation
\begin{equation}\label{incidence}
\omega_a(z)=X_{\mu}(z)\,\Gamma^{\mu}_{ab}\,\lambda^b(z).
\end{equation}
We define an ambitwistor coordinate as ${\cal Z}=(\omega_a, \lambda^a)$ and, using the incidence relations (\ref{incidence}), the action (\ref{XP}) may be written as
\begin{equation}\label{S2}
S= \int_{\Sigma}{\cal Z}\hspace{-.06cm}\cdot\hspace{-.06cm}\bar{\partial}{\cal Z}+\mu T.
\end{equation}
where the $P^2(z)=0$ constraint is solved automatically by (\ref{P}) and we have adopted the notation ${\cal Z}\hspace{-.06cm}\cdot\hspace{-.06cm}\bar{\partial}{\cal Z}:= \frac{1}{2}(\omega_a\bar{\partial}\lambda^a-\lambda^a\bar{\partial}\omega_a)$. The stress tensor is $T={\cal Z}\hspace{-.06cm}\cdot\hspace{-.06cm}\partial{\cal Z}$. As written, the target space of this sigma model is not the space of null lines but is much larger. A constraint must be introduced to reduce the target space of the naive the embedding ${\cal Z}:\Sigma\rightarrow\mathbb{C}^{2(2d-4)}$ to the  ($2d-3$ dimensional) physical space appropriate for a null line.

\subsection{The Constraint}

In this section we review the construction of the constraint in ten dimensions (see \cite{Cederwall:1992bi,Berkovits:1990yr} for details and the Appendix for a brief discussion\footnote{More on the Octonions may be found in \cite{Baez:2001dm}.}). The key idea of this section is the constraint (\ref{G2}) which generates gauge transformations that preserve $P_{\mu}(z)$ and the eager reader may safely skip the details of this section on a first reading. We shall take $d=10$ and focus on the bosonic sector of the theory. The supersymmetric extension is discussed briefly in the Appendix and we anticipate the inclusion of the additional sectors required by supersymmetry will be straightforward.

It is useful to package the sixteen $\lambda^a(z)$ into a two component spinor of $SL(2;\mathbb{O})$, which we write as
$\lambda_I(z)=\left(	\lambda^+(z) , \lambda^-(z)\right)$. Following \cite{Berkovits:1990yr}, we set $\lambda^a(z)=(\lambda^+_A(z),\lambda^-_A(z))$, where $A=1,2,...8$ and so
$$
\lambda^+(z)=\sum_{A=1}^8\lambda^+_A(z)\,e_A,	\qquad	\lambda^-(z)=\sum_{A=1}^8\lambda^-_A(z)\,e_A.
$$
Here $e_8=1$ and the $e_i=-\bar{e}_i$, where $i=1,2,...,7$, is a representation of the Octonions. Similarly, the gamma matrices may be written as $SL(2;\mathbb{O})$ matrices $\Gamma^{\mu}_{IJ}$. This division of the sixteen components of $\lambda^a$ into the $8+8$ components of $\lambda^{\pm}$ clearly breaks manifest Lorentz covariance but we shall see that it is ultimately possible to work in terms of covariant objects only at the cost of introducing a redundancy in the description of the constraints.

The $\Gamma^{\mu}_{IJ}$ satisfy the Clifford algebra relation and care must be taken to keep track of the order of operations as the Octonions are not associative. It will be helpful to introduce $\Gamma^A=(\Gamma^i,\Gamma^8)$, where $i=1,2,..7$, so that $\Gamma^{\mu}=(\Gamma^+,\Gamma^A,\Gamma^-)$ where $2\Gamma^{\pm}=\Gamma^0\pm\Gamma^9$. We can write the momentum in $SL(2;\mathbb{O})$ notation as $P_{IJ}=\Gamma_{IJ}^{\mu}P_{\mu}=\lambda_I\bar{\lambda}_J$ where
$$
P_{IJ}=\left(\begin{array}{cc} \sqrt{2}P^+ & P^Ae_A \\  P^A\bar{e}_A & \sqrt{2}P^-  \end{array}\right),
$$
where $I=1,2$ and $\sqrt{2}P^+=|\lambda^-|^2$, $P^Ae_A=\lambda^+\bar{\lambda}^-$, and $\sqrt{2}P^- =|\lambda^+|^2$. It is clear that $P^2=\det(P_{IJ})=0$.

We are interested in transformations of the $\lambda^a(z)$ that preserve the momentum $P_{\mu}(z)$ (\ref{P}). The transformations $\delta\lambda^+=\{\varepsilon^iU^+_i,\lambda^+\}=\lambda^+O^+(\varepsilon)$ and $\delta\bar{\lambda}^+=\{\bar{U}_i,\lambda^+\}=\bar{O}^+(\varepsilon)\bar{\lambda}^+$, where $O_i^+$ is some linear transformation on the spinor and $\varepsilon^i$ is  parameter, leaves $P_{++}=|\lambda_+|^2$ invariant if $U^+_i=-\bar{U}^+_i$. A similar argument follows for the $|\lambda^-|^2$ component, giving a generator $U_i=U_i^++U_i^-$. To leave $P^Ae_A=\lambda_1\bar{\lambda}_2$ invariant we require that $U_i^+$ and $U_i^-$ be related by $(\lambda^+O^+)\bar{\lambda^-}= \lambda^+(O^-\bar{\lambda}^-)$.  For $d=3,4,6$ the algebra is associative so $U^+_i$ and $U^-_i$ take the same form and we are able to write the generator in terms of $SL(2;\mathbb{K}_{d-2})$ spinors. In ten dimensions, the  algebra is not associative and it is not possible to write the generator in an $SL(2;\mathbb{O})$ covariant form \cite{Cederwall:1992bi}. 
 
A natural choice for the linear action on the spinor is $\lambda^+O^+(\varepsilon)=\lambda^+\varepsilon^je_j$ (Octonion multiplication on the right). The condition $(\lambda^+O^+)\bar{\lambda^-}= \lambda^+(O^-\bar{\lambda}^-)$ determines the corresponding transformation of $\lambda^-$, a modified Octonion multiplication from the left. Introducing conjugate variables $\omega_I$, one can find explicit expressions for the generators $U_i^{\pm}$.  For $d=3,4,6$ the algebra is associative and in these cases we can take\footnote{$\omega$ and $\lambda$ are conjugate so this is a rotation of the components of $\lambda$ where $\omega\sim \partial/\partial\lambda$.} the generator to have the form $J=U+\bar{U}$. Specifically,
 $$
 J=\frac{1}{2}\Big( \lambda^{\dagger I}\omega_I-\omega^{\dagger}_I\lambda^I\Big)
 $$
for $d=4$ which generates $U(1)$ and
$$
 J_i=\frac{1}{2}\Big( \lambda^Ie_i\bar{\omega}_I-\omega_Ie_i\bar{\lambda}^I\Big),	\qquad	i=1,2,3
 $$
 for $d=6$. Here $e_i$ is one of  the fundamental Quaternion units $\{i,j,k\}$ and the $J_i$ generate $SU(2)$, generalising the $U(1)$ in four dimensions. In general, one can show that the $P_{\mu}(z)$ are preserved by transformations generated by $J^i=U^i+\bar{U}^i$ where \cite{Berkovits:1990yr}
$$
U^i=\lambda^+e_i\bar{\omega}_++\frac{(\lambda^-\bar{\lambda}^+)}{|\lambda^+|^2}(\lambda^+e_j)\bar{\omega}_-.
$$
The commutator of two $J^i$ gives rise to structure functions \cite{Berkovits:1990yr} $h_{ij}{}^k(\lambda)+\bar{h}_{ij}{}^k(\lambda)$
$$
h_{ij}{}^k(\lambda)=-i\frac{\bar{\lambda}^+}{2|\lambda^+|^2}\Big( (\lambda^+e_j)e_i-(\lambda^+e_i)e_j \Big)e_k.
$$
The Quaternions are associative and so, in six dimensions, the structure functions become $-i[e_i,e_j]e_k=i\varepsilon_{ijk}$, the structure constants of $SU(2)$. In ten dimensions, unlike the other cases, the failure of associativity means that the algebra is necessarily field dependent and is not a conventional Lie algebra, but is a \emph{soft} algebra \cite{Sohnius:1982rs}. We shall see this field-dependence of the algebra appear explicitly in models later on where it will lead to a gauge algebra that only closes on-shell. Thus, the requirement that we ultimately introduce anti-fields to quantise the theory can be traced back to the failure of the Octonions to be associative.

The symmetry generator $J^i$ in ten dimensions breaks manifest Lorentz invariance. Writing  $G^I=(G_+,G_-)$ where $G_{\pm}=\lambda^i_{\pm}J^i$, we have a generator $G^I$ that transforms as an $SL(2;\mathbb{O})$ spinor \cite{Cederwall:1992bi}. For example, in four dimensions we may take $G^J= \lambda^I\lambda^J\bar{\omega}_I-\omega_IP^{JI}$, where we have written $\lambda^J\bar{\lambda}^I=P^{JI}$. Unlike the $J^i$, this form of the constraint is the same in $d=3,4,6$ and $10$. In ten dimensions the constraint may be written as the $SO(9,1)$ spinor
\begin{equation}\label{G2}
G^a=(\lambda^c\Gamma^{\mu}_{cd}\lambda^d)\Gamma_{\mu}^{ab}\omega_b-2\lambda^a\lambda^b\omega_b,
\end{equation}
where an overall normalisation has been chosen for convenience. The price to be paid in working with a manifestly Lorentz covariant formalism is that not all of the sixteen $G^a$ can be linearly independent and so the symmetry generated by the $G^a$ is reducible.

The commutator of two generators is
\begin{equation}\label{algebra1}
[G^a,G^b]=f^{ab}_c(\lambda)G^c,
\end{equation}
where $f^{ab}_c(\lambda)$ are functions of $\lambda^a(z)$. It is worth noting that, in any dimension, $f^{ab}_c(\lambda)$ are functions of $\lambda^a(z)$. This is due to the fact that the extra factor of $\lambda^a$ included in going from the generators $U^i$ to $G^a$ means that, whilst the $h_{ij}{}^k$ may or may not be constant, depending on the dimension under consideration, dimensional analysis alone requires that the $f^{ab}_c$ will be functions of $\lambda^a(z)$.

\subsection{The Sigma Model and its Symmetries}

Incorporating the constraint (\ref{G2}) into the action (\ref{S2}) leads to the constrained action
\begin{equation}\label{S3}
S= \int_{\Sigma}  {\cal Z}\hspace{-.06cm}\cdot\hspace{-.06cm}\bar{\partial}{\cal Z}+\mu T+e_aG^a,
\end{equation}
where we have included a Lagrange multiplier $e_a$ to impose the constraint $G^a=0$. The $2d-4$ functions $G^a$ are not linearly independent but are related by the $d$ expressions
$$
G^aZ_a^{\mu}=0,
$$
where $Z^{\mu}_a(z)\equiv\Gamma^{\mu}_{ab}\lambda^b(z)$. The transformation of $\lambda^a(z)$ generated by $G^a(z)$ preserves $P_{\mu}(z)$. It is not hard to see that this requires $(\delta \lambda^a)\Gamma_{ab}^{\mu}\lambda^b=0$ and so $\delta\lambda^a$ and hence $G^a$ is in the kernel of $Z_a^{\mu}$. This is a direct result of (\ref{identity}). In turn $Z^{\mu}_a$, treated as a map from the space of spinors $\lambda$ to Minkowski space, has a non-empty kernel as  a result of the identity $Z^{\mu}_aP_{\mu}=0$. There are thus only $d-3$ independent constraints. The space of null lines is $2d-3$ dimensional. Subject to the constraint $G^a=0$, the ambitwistors have $Z^I$ have $2(2d-4)-(d-3)=3d-5$ components. The symmetry $\delta{\cal Z}^I=C{\cal Z}^I$ removes one degree of freedom and the remaining $d-2$ non-physical degrees of freedom are removed by the gauge symmetry generated by $G^a$ which we now review. 

As is clear from the construction outlined in the previous section, $G^a$ does not act symmetrically on $\lambda^a(z)$ and $\omega_a(z)$ and so it is useful to work in terms of these spinor fields rather than the twistor ${\cal Z}(z)$.  These fields transform as \cite{Berkovits:1990yc}
$$
\delta\lambda^a=\Big(-(\lambda^c\Gamma^{\mu}_{cd}\lambda^d)\Gamma_{\mu}^{ab}+2\lambda^a\lambda^b\Big)\varepsilon_b,	\qquad	\delta\omega_a=\Big(2\Gamma^{\mu}_{ac}\lambda^c(\Gamma^{de}_{\mu}\omega_e)-2\delta^d_a(\lambda^e\omega_e)-2\lambda^d\omega_a\Big)\varepsilon_d,
$$
where $\varepsilon_a$ depends on the worldsheet coordinates. It will be useful to introduce
\begin{equation}\label{xi}
\xi^{ab}\equiv-(\lambda^c\Gamma^{\mu}_{cd}\lambda^d)\Gamma_{\mu}^{ab}+2\lambda^a\lambda^b,	\qquad	
\tilde{\xi}_a{}^b\equiv 2\Gamma^{\mu}_{ac}\lambda^c(\Gamma^{be}_{\mu}\omega_e)-2\delta^b_a(\lambda^e\omega_e)-2\lambda^b\omega_a
\end{equation}
so that $G^a=\xi^{ab}\omega_b$, $\delta\lambda^a=\xi^{ab}\varepsilon_a$ and $\delta\omega_a=\tilde{\xi}_a{}^b\varepsilon_b$. In addition there are conformal transformations generated by the stress tensor $T(z)$. It is useful to write the generators of infinitesimal conformal and gauge transformations as 
$$
\mathbf{G}(\varepsilon)=\oint\rd z\;\varepsilon_a(z)G^a(z),	\qquad	\mathbf{T}(\nu)=\oint\rd z\;v(z)T(z),
$$
respectively, where $v(z)$ is a weight $(-1,0)$ field (a worldsheet vector), $\varepsilon_a(z)$ is a weight $(-1/2,0)$ field and $T(z)$ is the stress tensor. The contour is implicitly assumed to encircle the point on $\Sigma$ where the operator is inserted. The algebra of the $G^a$'s is $[\mathbf{G}(\varepsilon),\mathbf{G}(\breve{\varepsilon})]=\mathbf{G}(\tilde{\varepsilon})$ where
$\tilde{\varepsilon}_{c}(z)=4\delta_c^{[a}\lambda^{b]}(z)\varepsilon_{a}(z)\breve{\varepsilon}_{b}(z)$ and may be written as (\ref{algebra1}) from which we see that the constraints are first class. We note that the algebra is not a Lie algebra, but has structure functions
\begin{equation}\label{f}
f_a^{bc}(\lambda)=-4\delta_a^{[b}\lambda^{c]}(z).
\end{equation}
Such algebras have been studied in many contexts \cite{Sohnius:1982rs} and are sometimes referred to as \emph{soft} gauge algebras\footnote{As opposed to a `hard' gauge algebra that would feature structure constants.}. This $\lambda^a$-dependence will play a crucial role in what follows and, as shown in \cite{Cederwall:1992bi,Berkovits:1990yr}, the field dependence in $f^{ab}_c(\lambda)$ can be traced back to the failure of associativity of the Octonions. If we adopt the notation
$$
[\nu_i,\nu_j]=\nu_i\partial \nu_j-\nu_j\partial \nu_i,		\qquad	[\nu_i,\varepsilon_{j}]_a=\frac{1}{2}\varepsilon_{ja}\partial\nu_i-\nu_i\partial \varepsilon_{ja},	\qquad	[\varepsilon_{i},\varepsilon_{j}]_a=f_a^{bc}(\lambda)\varepsilon_{ib}\varepsilon_{jc}.
$$
The full gauge algebra of the theory may then be written as
$$
[\mathbf{T}(v_i),\mathbf{T}v_j)]=-\mathbf{T}([v_i,v_j]),	\qquad	[\mathbf{T}(v_i),\mathbf{G}(\varepsilon_j)]=-\mathbf{G}([v_i,\varepsilon_j]),
$$
\begin{equation}\label{algebra}
[\mathbf{G}(\varepsilon_i),\mathbf{G}(\varepsilon_j)]=-\mathbf{G}([\varepsilon_i,\varepsilon_j]).
\end{equation}

\subsection{Transformation of $e_a$}

The Lagrange multiplier $e_a(z)$ is a weight $(-1/2,1)$ bosonic field and invariance of the action (\ref{S3}) requires it to transform as a connection
$$
\delta e_a=-\bar{\partial}\varepsilon_a-4\delta_a^{[b}\lambda^{c]}\varepsilon_ce_b,
$$
and the algebra of these transformations closes on shell
\begin{equation}\label{newalg}
[\delta_{\varepsilon_1},\delta_{\varepsilon_2}]e_a=\delta_{\varepsilon_3}e_a+4\delta_a^{[b}\delta^{c]}_d\frac{\delta S}{\delta \omega_d}\varepsilon_b\tilde{\varepsilon_c},
\end{equation}
where $\varepsilon_3=-4\delta_a^{[b}\lambda^{c]}\varepsilon_{2b}\varepsilon_{1c}$ and
$$
\frac{\delta S}{\delta \omega_a}=\bar{\partial}\lambda^a+\xi^{ab}e_b,
$$
gives the $\lambda^a$ equation of motion. The algebra may be written as
$$
[G^a,G^b]=f_{c}^{ab}G^c+4q\delta_c^{[a}\delta^{b]}_d\frac{\delta S}{\delta \omega_d},
$$
where $q=1$ for $e_a(z)$ and $q=0$ for other fields. The fact that the algebra (\ref{newalg}) is open on $e_a$ (i.e. only closes up to equations of motion) will have repercussions on how we deal with defining path integrals in the theory.

This is not the most general transformation of $e_a$ that is a symmetry of the action, we may also add a shift term proportional\footnote{The sign is chosen for later convenience.} to $Z^{\mu}_a$, since $Z^{\mu}_aG^a=0$ identically by virtue of the identity (\ref{identity}), and so we have
\begin{equation}\label{e2}
\delta e_a=-\bar{\partial}\varepsilon_a-4\delta_a^{[b}\lambda^{c]}\varepsilon_ce_b-Z^{\mu}_a\varepsilon_{\mu}.
\end{equation}
This expresses the reducibility of the gauge symmetry and the identity $Z^{\mu}_aP_{\mu}=0$ means that only $d-1$ of the $\varepsilon_{\mu}$ are independent.

This reducibility is easy to understand at the level of the field transformations. Consider the gauge transformations with parameter $\varepsilon_a=Z^{\mu}_a\varepsilon_{\mu}$. It is not hard to see that $\delta\lambda^a=\xi^{ab}Z^{\mu}_b\varepsilon_{\mu}=0$, for any $\varepsilon_{\mu}(z)$. The variation of the action under a gauge transformation with parameter $Z^{\mu}_a\varepsilon_{\mu}$ is
$$
\delta S= \int \delta\omega_a\left(\bar{\partial}\lambda^a+\xi^{ab}e_b\right)+\delta e_aG^a.
$$
On-shell, where $\bar{\partial}\lambda^a+\xi^{ab}e_b=0$, this implies that $\delta e_a=0$ and so in the path integral, when we quotient out by the action of the gauge group, we want to exclude such transformations from consideration. It is also clear that the shift $\varepsilon_{\mu}\rightarrow \varepsilon_{\mu}+P_{\mu}\varepsilon$ has no effect on the above argument and so we want to think of the $\varepsilon_{\mu}$ as only being defined up to the addition of $P_{\mu}\varepsilon$. The algebra closes on-shell and so the quantisation of this theory will require the full BV treatment, an issue we will deal with in section four.

\section{The BRST Charge}

We introduce a BRST charge $Q$ which acts on the space of fields (except $e_a(z)$). This construction will be algebraic and will have nothing to say about the action of the theory in question. If we neglect the $e_a(z)$ field, the algebra closes off shell on the fields and we can follow standard procedure to construct suitable BRST charge. The fact that the algebra only closes on-shell on the $e_a(z)$ field will be dealt with in section four.

\subsection{A sketch of gauge-fixing}

To streamline the presentation we will ignore the conformal transformations in what follows, effectively setting $b(z)$ and $c(z)$ to zero. It is straightforward to accommodate the conformal ghost sector later.

\vspace{.3cm}
\noindent\emph{Ghosts}
\vspace{.2cm}

\noindent We start with the Lagrangian ${\cal L}_0+e_aG^a$, where ${\cal L}_0=   {\cal Z}\hspace{-.06cm}\cdot\hspace{-.06cm}\bar{\partial}{\cal Z}$. This has symmetry
$$
\delta e_a=\bar{\partial}\varepsilon_a+f_a^{bc} \varepsilon_b e_c,	\qquad	\delta \lambda^a=\xi^{ab} \varepsilon_b,	\qquad	\delta\omega_a=\tilde{\xi}_a{}^b\varepsilon_b,
$$
where $\xi^{ab}$ and $\tilde{\xi}_a{}^b$ are given by (\ref{xi}). We introduce ghosts $(\eta_a,\rho^a)$ and the BRST charge
\begin{equation}\label{firstQ}
Q=\oint\rd z\;\eta_a\Big(G^a+\frac{1}{2}f^{ab}_c\eta_b\rho^c\Big).
\end{equation}
$\eta_a(z)$ and $\rho^a(z)$ are of conformal weight $-1/2$ and $3/2$ respectively and satisfy the anti-commutation relations $\{\rho^a(z),\eta_b(w)\}=\delta^a_b\delta(z-w)$. This BRST charge tells us how to augment the constraint $G^a$ so that it also acts on the ghosts and we introduce
$$
H^a\equiv\{Q,\rho^a\}=G^a+f^{ab}_c\eta_b\rho^c.
$$
Before we move on to consider the reducibility of these generators, we briefly consider the ghost-modified generator $H^a$. It was already observed, in a slightly different context in \cite{Cederwall:1992bi}, that the algebra of the $H^a$ does not close without the introduction of additional generators.
$$
[H^a,H^b]=f^{ab}_cH^c+f^{abc}_dJ_c{}^d,	\qquad	[H^a,J_b{}^c]=f^{ad}_bJ_d{}^c-f^{ac}_dJ_b{}^d,	
$$
$$
[J_a{}^b,J_c{}^d]=\delta_c^bJ_a{}^d-\delta_a^dJ_c{}^b
$$
where $f^{ab}_c$ is given by (\ref{f}) and  $f^{abc}_d=4\delta^{[a}_d\xi^{b]c}$ with $\xi^{ab}$ given by (\ref{xi}). We have introduced the generators $J_a{}^b=\eta_a\rho^b$ which simply exchange the ghost fields amongst themselves. The enlargement of the soft algebra by the currents $J_a{}^b(z)$ may play an interesting role but we do not consider it further here.

There is an obstruction to setting the fields $e_a(z)$ to zero globally. The best we can do is to fix
\begin{equation}\label{gf1}
F_a:=e_a(z)-\sum_r s_a^r(z) \tilde{\mu}_r=0,
\end{equation}
where $\tilde{\mu}_r$ is a basis for the moduli space of these fields and $s_a^r(z)$ are worldsheet fields, which we take to transform under the BRST transformation as $\delta s_a^r(z)=m_a^r(z)$, where $\delta m_a^r(z)=0$. We then introduce the gauge-fixing fermion
$$ 
\psi=\int_{\Sigma}\rd^2 z\;\rho^a(z) \left(e_a(z)- \sum_{r}s_a^r(z) \tilde{\mu}_r \right).
$$
The (non-minimal) action is then given by the action
$$
S=\delta\psi+\int_{\Sigma}{\cal Z}\hspace{-.06cm}\cdot\hspace{-.06cm}\bar{\partial}{\cal Z}+b\bar{\partial}c+F_a\pi^a,
$$
where we have introduced an auxiliary field $\pi_a$ in the gauge-fixing term and $\delta\psi$ denotes the BRST variation of $\psi$. It is useful to define an inner product $\langle\;,\;\rangle$ given by integration over $\Sigma$, for example
$$
\langle H^a,\tilde{\mu}_r\rangle=\int_{\Sigma}\rd^2z\,H^a(z)\, \tilde{\mu}_r(z),	\qquad	\langle b,\mu_m\rangle=\int_{\Sigma}\rd^2z\,b(z)\, \mu_m(z),
$$
where $\mu_m$ ($m=1,2,...,n+3g-3$) are Beltrami differentials defined by $\mu_m=\partial\mu/\partial\tau^m$ and $\tau^m$ ($m=1,2,...n+3g-3$) are local holomorphic coordinates on the moduli space of an n-puntured genus $g$ Riemann surface. Using standard techniques and including the contribution from the conformal ghosts we find that the correlation functions of observables are naively given by
\begin{equation}\label{action1}
\langle {\cal V}_1(z_1)...{\cal V}_n(z_n)\rangle=\int_{\Gamma_n}\left\langle\prod_m\langle b,\mu_m\rangle\;\prod_{a,r}\langle\rho^a,\tilde{\mu}_r\rangle\;\delta\Big(\langle H^a,\mu_r\rangle\Big)\; {\cal V}_1(z_1)...{\cal V}_n(z_n)\right\rangle,
\end{equation}
where the correlation function under the integral is computed using a path integral with the action
\begin{equation}\label{S4}
S=\int_{\Sigma}{\cal Z}\hspace{-.06cm}\cdot\hspace{-.06cm}\bar{\partial}{\cal Z}+\rho^a\bar{\partial}\eta_a+b\bar{\partial}c.
\end{equation}
and ${\cal V}_i(z_i)$ are physical operators inserted at the point $z_i$, corresponding to some observable (i.e. in the cohomology of $Q$). The expression (\ref{action1}) cannot be the full story as we have not taken into account the reducibility of the gauge symmetry and the BRST operator (\ref{firstQ}) is not quite right. The cycle of integration $\Gamma_n$ is over an appropriate space and will be discussed briefly in section 4.5. A complaint could be levelled at the above expression in that the action is not BRST invariant. This is due to the fact that $\delta^2 e_a$ is only weakly zero, i.e. it is zero up to terms proportional to the equations of motion and so $\delta^2\psi\neq 0.$ This issue is easily dealt with within the framework of BV quantisation \cite{Batalin:1984jr,Batalin:1981jr} and we shall return to this issue in section 4.

\vspace{.3cm}
\noindent\emph{Ghosts for ghosts}
\vspace{.2cm}

\noindent The action (\ref{S4}) has the additional fermionic symmetry generated by $H^{\mu}=Z^{\mu}_a\rho^a$ and given by
\begin{equation}\label{trans}
\delta'\lambda^a=0,	\qquad	\delta'\omega_a=\Gamma^{\mu}_{ab}\rho^b \varepsilon_{\mu},	\qquad	\delta'\eta_a=Z^{\mu}_a\varepsilon_{\mu}.
\end{equation}
We add to the action the term $e_{\mu}H^{\mu}$, mirroring $e_aG^a$, and find the above transformations are symmetry of the action (\ref{S4}) if $e_{\mu}(z)$ transforms as
$$
\delta' e_{\mu}=\bar{\partial}\varepsilon_{\mu},
$$
which is consistent with $[H^{\mu},H^{\nu}]=0$. Note that $\varepsilon_{\mu}(z)$ is a grassmann parameter. The idea that $H^{\mu}=0$ should be fixed by a Lagrange multiplier is reminiscent of the condition that the `$b$-ghosts' annihilate the physical state $b|\Psi\rangle=0$. In this case it is tempting to require that $\rho^a|\Psi\rangle=0$; however, this is too many conditions on $|\Psi\rangle$ as not all of the $\rho^a$ are independent of each other. Instead the relevant condition is $\rho^a|\Psi\rangle=0$, suplimented with $Z^{\mu}_a\rho^a=0$. The condition on $Z^{\mu}_a\rho^a$ removes $d$ of the $\rho^a$, leaving $d-4$ independent degrees of freedom so that $\rho^a|\Psi\rangle=0$ imposes $d-4$ conditions on $|\Psi\rangle$.

We gauge fix the symmetry generated by (\ref{trans}) as above by introducing ghosts $(\rho^{\mu},\eta_{\mu})$. $\rho^{\mu}$ and $\eta_{\mu}$ have Bose statistics and are of weight $2$ and $-1$ respectively. They satisfy the commutation relation $[\rho^{\mu}(z),\eta_{\nu}(w)]=\delta^{\mu}_{\nu}\delta(z-w)$. We introduce the BRST charge for the transformations (\ref{trans})
$$
Q'=\oint\rd z\;\eta_{\mu}(z)H^{\mu}(z),
$$
and a gauge-fixing fermion
$$ 
\psi'=\int_{\Sigma}\rd^2 z\;\rho^{\mu}(z) \left(e_{\mu}(z)-\sum_{\dot{r}}s_{\mu}^{\dot{r}}(z)\tilde{\mu}_{\dot{r}}\right),
$$
where the $\tilde{\mu}_{\dot{r}}$ are a basis for the moduli space of the $e_{\mu}(z)$ fields. $\delta\psi'$ will provide a kinetic term for the new ghosts $\rho^{\mu}\bar{\partial}\eta_{\mu}$ in the gauge-fixed action.

\vspace{.3cm}
\noindent\emph{Ghosts for ghosts for ghosts}
\vspace{.2cm}

\noindent If we take as a starting point the Lagrangian ${\cal L}_0+\rho^a\bar{\partial}\eta_a+\rho^{\mu}\bar{\partial}\eta_{\mu}$ we find the action has the residual (bosonic) symmetry
$$
\delta''\lambda^a=0,	\qquad	\delta''\omega_a=2 \varepsilon \rho^{\mu}\Gamma_{\mu ab}\lambda^b,	\qquad	\delta''\eta_{\mu}=\varepsilon Z_{\mu}.
$$
In the same way as above, we introduce a constraint on the ghosts $\rho^{\mu}$, only $d-1$ of which are independent
$$
H=\rho^{\mu}(\lambda^a\Gamma_{\mu ab}\lambda^b)=\rho^{\mu}Z_{\mu}=0.
$$
This ensures that the condition $\rho^{\mu}|\Psi\rangle=0$ only places $d-1$ constraints on physical states. Thus, the conditions $H^a=0$, $H^{\mu}=0$, and $H=0$ ensure that $\rho^a|\Psi\rangle=0$ places exactly $d-3$ constraints on the state. We introduce a fermionic ghost system $(\rho,\eta)$ where $\rho$ and $\eta$ have conformal weight $3$ and $-2$ respectively and obey the anti-commutation relation $\{\rho(z),\eta(w)\}=\delta(z-w)$. The field transformations are generated by the BRST charge
$$
Q''=\oint\rd z\;\eta(z) H(z),
$$
and $\{Q'',\rho(z)\}=H(z)$. The Lagrangian
$$
{\cal L}={\cal L}_0 +\rho^a\bar{\partial}\eta_a+\rho^{\mu}\bar{\partial}\eta_{\mu}+eH,
$$
is invariant under the symmetry generated by $H$ if the Lagrange multiplier $e(z)$ transforms as $\delta'' e=\bar{\partial}\varepsilon$. Introducing the gauge-fixing fermion
$$ 
\psi''=\int_{\Sigma}\rho(z)  \left(e(z)-\sum_{\ddot{r}}s^{\ddot{r}}(z)\tilde{\mu}_{\ddot{r}}\right),
$$
the gauge-fixed action now also includes the kinetic term $\rho\,\bar{\partial}\eta$ for these new ghosts.

\subsection{The BRST Charge in detail}

The problem with the preceding discussion is that it neglects the possible effect of terms in the BRST charge that may involve interaction terms between the different ghost sectors. It is not hard to see that the the total charge
$$
\oint\rd z \;c\,{\cal Z}\hspace{-.06cm}\cdot\hspace{-.06cm}\partial{\cal Z}+\eta_a\left(G^a+\frac{1}{2}f^{ab}_c\eta_b\rho^c\right)+\rho^aZ_a^{\mu}\eta_{\mu}+\rho^{\mu} P_{\mu}\eta=\oint\rd z\;c\,{\cal Z}\hspace{-.06cm}\cdot\hspace{-.06cm}\partial{\cal Z}+\eta_aH^a+\eta_{\mu}H^{\mu}+\eta H,
$$
does not square to zero. The crude sketch above gives a feel for the role of the additional ghost sectors but neglects important details. We now turn to a more careful discussion of gauge-fixing and the construction of the BRST charge. We will take the BRST charge to have the form
$$
Q=\oint\rd z\; c\left(T+\frac{1}{2}T_{gh}\right)+\eta_a\left(G^a+\frac{1}{2}f^{ab}_c\eta_b\rho^c\right)+\rho^aZ_a^{\mu}\eta_{\mu}+\rho^{\mu} Z_{\mu}\eta+...,
$$
where the $+...$ terms denote terms required to ensure that $Q^2=\frac{1}{2}\{Q,Q\}=0$, which we seek to determine. Here $\{\cdot,\cdot\}$ is a Poisson bracket\footnote{Our focus will be on the construction of the classical gauge-fixed action.} given by
$$
\{A,B\}= \sum_I\int_{\Sigma}\rd^2 z\left(\frac{\delta A}{\delta \phi^I(z)} \frac{\delta B}{\delta \chi_I(z)}- \frac{\delta A}{\delta \chi_I(z)} \frac{\delta B}{\delta \phi^I(z)}\right),
$$
where $\phi^I$ denotes the fields $(\lambda^a,c,\eta_a,\eta_{\mu},\eta)$ and $\chi_I$ are the conjugate fields $(\omega_a,b,\rho^a,\rho^{\mu},\rho)$ which may be thought of as the functional derivatives $(\delta/\delta \lambda^a,\delta/\delta c,\delta/\delta\eta_a,\delta/\delta \eta_{\mu},\delta/\delta\eta)$. The stress tensor $T(z)$ includes contributions from all ghost sectors, with the exception of the conformal $(b,c)$ system, which has stress tensor $T_{gh}(z)$. A suitable ansatz for the BRST charge is
\begin{equation}\label{Q1}
Q=\oint\rd z\; c\left(T+\frac{1}{2}T_{gh}\right)+\sum_{p=0}^3Q_p,
\end{equation}
where $T$ is now the stress tensor for all sectors except for the $(b,c)$ conformal ghosts
\begin{equation}\label{T}
T=-\frac{1}{2}\Big(\omega_a\partial \lambda^a-\lambda^a\partial\omega_a \Big)+ T_{3/2}+T_2+T_3,
\end{equation}
where
$$
T_{3/2}=\frac{1}{2}\eta_a\partial \rho^a-\frac{3}{2}\rho^a\partial\eta_a,	\qquad	T_2=-\eta_{\mu}\partial \rho^{\mu}-2\rho^{\mu}\partial\eta_{\mu},	\qquad	T_3=2\eta\partial \rho-3\rho\partial\eta.
$$
The conformal ghost stress tensor takes the usual form $T_{\text{gh}}=c\partial b-2b\partial c$. This issue is largely orthogonal to the role played by the conformal ghosts and, to streamline the presentation, we shall suppress all mention of the conformal ghosts until the end. The constraint $G^a$ is linear in $\omega_a$ and so we only need consider an ansatz for $Q$ that is linear in the $\rho$-ghosts \cite{Ferraro:1992ek}.  The remaining contributions to the BRST charge are
\begin{eqnarray}
Q_0&=&\oint\rd z\; \eta_aG^a,\nonumber\\
Q_1&=&\oint\rd z\Big( \rho^aZ_a^{\mu}\eta_{\mu}+\frac{1}{2}f^{ab}_c\eta_a\eta_b\rho^c\Big),\nonumber\\
Q_2&=&\oint\rd z\Big( \rho^{\mu}Z_{\mu}\eta-{\cal C}^{a\mu}_{\nu}\eta_a\eta_{\mu}\rho^{\nu}+\frac{1}{6}{\cal M}_{\mu}^{abc}\eta_a\eta_b\eta_c\rho^{\mu}\Big),\nonumber\\
Q_3&=&\oint\rd z\Big( \frac{1}{2}{\cal N}^{\mu\nu}\eta_{\mu}\eta_{\nu}\rho-{\cal C}^a\eta_a\eta\rho+\frac{1}{2}{\cal M}^{ab\mu}\eta_a\eta_b\eta_{\mu}\rho\Big).
\end{eqnarray}
The structure functions were found previously $f^{ab}_c=4\delta^{[a}_c\lambda^{b]}$ and the reducibility factors are $Z^{\mu}_a=\Gamma_{ab}^{\mu}\lambda^b$ and $Z_{\mu}=\lambda^a\Gamma_{ab\mu}\lambda^b$. The condition $Q^2=0$ places restrictions on the functions ${\cal C}^{a\mu}_{\nu}$, ${\cal C}^a$, ${\cal N}^{\mu\nu}$, ${\cal M}_{\mu}^{abc}$, and ${\cal M}^{ab\mu}$. At the classical level, these restrictions take the form of differential equations in $\lambda^a$ 
\begin{eqnarray}\label{de}
\xi^{ad}f^{bc}_d&=&\xi^{bd}\partial_d\xi^{ac}-\xi^{cd}\partial_d\xi^{ab},\nonumber\\
\xi^{ac}\partial_cZ^{\mu}_b+Z^{\mu}_cf^{ca}_b&=&{\cal C}^{a\mu}_{\nu}Z^{\nu}_b,\nonumber\\
\xi^{[a|d}\partial_df^{|bc]}_e&=&f_d^{[ab}f^{c]d}_e-\frac{2}{3}{\cal M}^{abc}_{\mu}Z^{\mu}_{e},\nonumber\\
{\cal C}^{a\mu}_{\nu}Z^{\lambda}_{a}+{\cal C}^{a\lambda}_{\nu}Z^{\mu}_{a}&=&{\cal N}^{\mu\lambda}Z_{\nu},\nonumber\\
\xi^{[a|c}\partial_c{\cal C}^{|b]\mu}_{\nu}&=&-\frac{1}{2}f^{ab}_c{\cal C}^{c\mu}_{\nu}-{\cal C}^{[a|\mu}_{\lambda}{\cal C}^{|b]\lambda}_{\nu}+{\cal M}^{ab\mu}Z_{\mu}+{\cal M}^{abc}_{\mu}Z^{\mu}_c,\nonumber\\
{\cal C}^aZ^{\mu}_a&=&{\cal N}^{\mu\nu}Z_{\nu},\nonumber\\
\xi^{[a|c}\partial_c{\cal C}^{|a]}&=&-\frac{1}{2}f_{c}^{ab}{\cal C}^c+{\cal C}^{[a}{\cal C}^{b]}+{\cal M}^{ab\mu}Z_{\mu}.
\end{eqnarray}
where $\partial_a=\partial/\partial\lambda^a$. The requirement that anomalies vanish place additional conditions on the functions and we shall discuss the vanishing of the conformal anomaly briefly below. The first equation in the list gives the structure functions (\ref{f}). We solve the other differential equations starting with the input data
$$
\xi^{ab}\equiv(\lambda^c\Gamma_{cd}^{\mu}\lambda^d)\Gamma_{\mu}^{ab}-2\lambda^a\lambda^b,	\qquad	Z^{\mu}_a\equiv\Gamma_{ab}^{\mu}\lambda^b,	\qquad	Z_{\mu}\equiv\eta_{\mu\nu}(\lambda^a\Gamma_{ab}^{\nu}\lambda^b).
$$
The equations (\ref{de}) may be solved to give
$$
f_a^{bc}=-4\lambda^{[b}\delta^{c]}_a,		\qquad	{\cal C}^{a\mu}_{\nu}=2\Gamma^{\mu ab}\Gamma_{\nu bc}\lambda^c, \qquad	{\cal C}^a=4\lambda^a,	\qquad	{\cal N}^{\mu\nu}=4\eta^{\mu\nu},
$$
\begin{equation}\label{constants}
 {\cal M}_{\mu}^{abc}=0, 	\qquad	{\cal M}^{ab\mu}=0.
\end{equation}
The classical BRST variation of all fields, with the exception of $e_a$, is then given by $\delta_Q\phi^i(z)=\{Q,\phi^i(z)\}$.

It will be useful to define the extended generators
\begin{equation}\label{G}
{\cal G}^a(z)=-\{Q,\rho^a(z)\},	\qquad	{\cal G}^{\mu}(z)=-\{Q,\rho^{\mu}(z)\},	\qquad	{\cal G}(z)=-\{Q,\rho(z)\}.
\end{equation}
Explicitly
$$
{\cal G}^a=G^a+f^{ab}_c\eta_b\rho^c-{\cal C}^{a\mu}_{\nu}\eta_{\mu}\rho^{\nu}-{\cal C}^a\eta\rho-\frac{1}{2}c\partial \rho^a+\frac{3}{2}\partial (c\rho^a),
$$
$$
{\cal G}^{\mu}=\rho^aZ^{\mu}_a-{\cal C}^{a\mu}_{\nu}\eta_a\rho^{\nu}+{\cal N}^{\mu\nu}\eta_{\nu}\rho-c\partial\rho^{\mu}+2\partial(c\rho^{\mu}),
$$
$$
{\cal G}=\rho^{\mu}Z_{\mu}+{\cal C}^a\eta_a\rho+5c\partial\rho-3\rho\partial c.
$$
${\cal G}^a$ clearly generalises $G^a$ to include the ghost sector. ${\cal G}^{\mu}$ and ${\cal G}$ may be thought of as nonlinear generalisations of the expressions $\rho^aZ^{\mu}_a$ and $\rho^{\mu}Z_{\mu}$ respectively.

Finally we note that the central charge of the theory, including the complete reducible ghost sector and the conformal ghosts may be straightforwardly calculated. We find that it vanishes if $d=26$. This is as expected from the traditional form of the theory presented in \cite{Mason:2013sva}. Details of this calculation may be found in the Appendix.

\section{The Master Action and Gauge Fixing}

We have noted at various points that the gauge algebra (BRST symmetry) only closes (is nilpotent) up to equations of motion. Given that the only physical field that exhibits this problem is $e_a(z)$, which we gauge fix, one could ask whether this really causes any problems in practice. For most considerations it is unlikely that this issue will cause any significant problems; however, a well-developed procedure to deal with such cases does exist \cite{Batalin:1984jr,Batalin:1981jr} and so, for completeness, we discuss this here.

\subsection{$e_a$ and its descendants}

The discussion so far has largely neglected $e_a(z)$. In particular, the BRST charge does not explicitly encode the gauge transformation of this field. We can generalise the BRST transformation of $e_a(z)$ to include the shift symmetry of (\ref{e2})
\begin{equation}\label{d2}
\delta e_a=-\bar{\partial}\eta_a+f_a^{bc} \eta_b e_c-e_{\mu}Z^{\mu}_a,
\end{equation}
where $e_{\mu}$ is a Grassmann odd field. We see that $\delta e_a=e_{\mu}Z^{\mu}_a+...$ acts as a shift symmetry which removes $d-1$ components\footnote{Since $Z^{\mu}_a(e_{\mu}+P_{\mu}\alpha)=Z^{\mu}_ae_{\mu}$ for any function $\alpha$.} of $e_a$. The BRST charge (\ref{Q1}) determines how the ghosts transform
$$
\delta \eta_a=Z^{\mu}_a\eta_{\mu}+\frac{1}{2}f_a^{bc}\eta_b\eta_c,	\quad	\delta\eta_{\mu}=Z_{\mu}\eta-{\cal C}^{a\nu}_{\mu}\eta_a\eta_{\nu},	\qquad	\delta\eta=\frac{1}{2}{\cal N}^{\mu\nu}\eta_{\mu}\eta_{\nu}-{\cal C}^a\eta_a\eta,
$$
with the coefficients given by (\ref{constants}). With a little work one can show that
$$
\delta^2e_a=-\Big(\bar{\partial}\lambda^b+\xi^{bc}e_c\Big)\Big( 2\eta_b\eta_a-\Gamma_{ab}^{\mu}\eta_{\mu} \Big)+Z^{\mu}_a\Big(-\delta e_{\mu}-\bar{\partial}\eta_{\mu}+{\cal C}^{b\nu}_{\mu}e_b\eta_{\nu}-{\cal C}^{b\nu}_{\mu}\eta_be_{\nu}\Big),
$$
where the identity
$$
\xi^{ac}\partial_cZ^{\mu}_b+Z^{\mu}_cf^{ca}_b={\cal C}^{a\mu}_{\nu}Z^{\nu}_b,
$$
from (\ref{de}) has been used to simplify the expression. If we generalise the BRST symmetry to include the transformations
 \begin{equation}\label{d1}
 \delta e_{\mu}\equiv -\bar{\partial}\eta_{\mu}+{\cal C}^{b\nu}_{\mu}e_b\eta_{\nu}-{\cal C}^{b\nu}_{\mu}\eta_be_{\nu}+eP_{\mu},
 \end{equation}
then $\delta^2e_a$ vanishes up to terms proportional to $\delta S/\delta\omega_a$. We are at liberty to include a shift symmetry $\delta e_{\mu}=...+eP_{\mu}$ to account for the reducibility at level one. Taking (\ref{d1}) as the BRST transformation of $e_{\mu}$, we now consider the second variation of $e_{\mu}$ and with a little work find
$$
\delta^2e_{\mu}=-\Big(\bar{\partial}\lambda^b+\xi^{bc}e_c\Big)\Big( 2 \Gamma^{\nu ad}\Gamma_{\mu db}\eta_a\eta_{\nu}-2Z_{\mu a}\eta \Big)+Z_{\mu}\Big( \delta e + \bar{\partial}\eta-{\cal C}^a\eta e_a-{\cal C}^a\eta_a e+{\cal N}^{\mu\nu}e_{\mu}\eta_{\nu}\Big),
$$
where the identity 
$$
\xi^{[a|c}\partial_c{\cal C}^{|b]\mu}_{\nu}+\frac{1}{2}f^{ab}_c{\cal C}^{c\mu}_{\nu}+{\cal C}^{[a|\mu}_{\lambda}{\cal C}^{|b]\lambda}_{\nu}=0,
$$
in (\ref{de}) has been used. The requirement that $\delta^2e_{\mu}$ vanishes up to terms proportional to $\delta S/\delta\omega_a$, fixes the BRST transformation of $e(z)$ and we generalise the BRST symmetry to include 
\begin{equation}\label{d3}
\delta e\equiv -\bar{\partial}\eta+{\cal C}^a\eta e_a-{\cal C}^a\eta_a e+{\cal N}^{\mu\nu}e_{\mu}\eta_{\nu}.
\end{equation}
With a bit of work one may show that, using the identities
$$
{\cal C}^{a\mu}_{\nu}Z^{\lambda}_{a}+{\cal C}^{a\lambda}_{\nu}Z^{\mu}_{a}={\cal N}^{\mu\lambda}Z_{\nu},	\qquad	{\cal C}^aZ^{\mu}_a={\cal N}^{\mu\nu}Z_{\nu},
$$
that $\delta e^2$ also vanishes up to terms proportional to $\delta S/\delta\omega_a$. In particular, we have
$$
\delta^2 e=-4(\bar{\partial}\lambda^a+\xi^{ab}e_b)\eta_a\eta.
$$
The fact that $\delta^2$ only vanishes up to equations of motion will require the widening of the space of fields to include anti-fields if we are to construct a BRST-invariant action. We shall define (\ref{d2}), (\ref{d1}) and (\ref{d3}) to be the BRST transformations of the fields $e_a(z)$, $e_{\mu}(z)$ and $e(z)$ respectively. We augment the BRST charge $Q$ to generate these transformations also. We will call this augmented BRST charge $\widehat{Q}$ and we shall see that the action of $\widehat{Q}$ on the space of fields is naturally incorporated into the BV framework.

\subsection{Deformations and Moduli}

The basic framework we have been exploring is that of a closed $n$-punctured Riemann surface $\Sigma$ with a bundle which has a soft algebra, generated by $G^a(z)$, associated with a natural $S^7$ action. There will be obstructions to setting the gauge fields to zero everywhere and upon gauge-fixing, the functional integrals over the fields $e_a(z)$, $e_{\mu}(z)$, $e(z)$ and $\mu(z)$  reduce to finite dimensional integrals over a moduli space which we shall denote by ${\cal E}$. The algebra (\ref{algebra}) suggests that we may think of ${\cal E}$ as a bundle over the moduli space of closed Riemann surfaces ${\cal M}$.

To integrate over this space we consider deformations of the underlying Riemann surface and the gauge bundle along the lines of \cite{DHoker:1988pdl}. Deformations in the moduli of $\Sigma$ are generated by the stress tensor $T(z)$ and, at genus zero, a basis for such deformations is given by translating the location of $n-3$ of the $n$ punctures. If we introduce a coordinate system $z_i$ in a small disc $\mathscr{D}_i$ around the $i$'th puncture, the moduli deformation is encoded in a worldsheet vector field $v(z_i)$ which gives $z_i\rightarrow z_i+v_m(z_i)\delta\tau^m$ with $\tau^m$ a holomorphic coordinate on ${\cal M}$. A basis of $n-3$ vectors will be denoted by $\vec{v}_m(z_i)=\Big(v_1(z_i),...,v_{n-3}(z_i)\Big)$ and it is natural to choose a basis such that we associate each component with a different puncture. In a small annulus around a puncture the vector field $v(z)$ is related to the Beltrami differential $\mu_m(z)=\partial_m\mu$ by $\bar{\partial}v_m=\mu_m$ where $\partial_m=\partial/\partial\tau^m$. And so we can encode the deformation as the charge
$$
\mathbf{T}(\vec{v}_m)=\int_{\Sigma}\rd^2z\,T(z)\mu_m(z)=\sum_{i=1}^n\oint_{\mathscr{D}_i}\rd z_i\,T(z_i)v_m(z_i),
$$
where we have used $\partial\Sigma=-\cup_{i=1}^n\partial\mathscr{D}_i$.  And similarly for the $b$-field
$$
\mathbf{b}(\vec{v}_m)\equiv \sum_{i=1}^n \oint_{\mathscr{D}_i}\rd z_i\;b(z_i)v_m(z_i).
$$
The $\vec{v}_m$ give rise to a basis for $T{\cal M}$ (see for example \cite{Zwiebach:1992ie} for details). We need a similar basis for the tangent to the fibres when integrating over the $e_a$ (and its descendants). The space of $e_a$ is the space of weight $(1,-1/2)$ worldsheet fields with bosonic statistics and so the moduli space for $e_a(z)$ is the space of such fields, modulo the gauge transformations (\ref{e2}). This space is familiar from the integration over worldsheet gravitini in the conventional superstring, except in this case we have a parity reversed field and so for each of the $2d-4$ $e_a(z)$ we have a copy of this $n-2$ dimensional space. Let $\{\tilde{\mu}_r\}$ be a basis for the $n-2$ dimensional tangent space to this moduli space. The gauge-fixed $e_a(z)$ may be written as
$$
e_a(z,\tau)=\sum_r s_a^r(z)\,\tilde{\mu}_r(z,\tau),
$$
just as in the standard gravitino case, except the $s_a^r(z)$ are worldsheet fields with the opposite statistics. As above, it is useful to introduce
\begin{equation}\label{G1}
\langle{\cal G}^a,\tilde{\mu}_r\rangle=\int_{\Sigma}\rd^2z\,{\cal G}^a(z)\,\tilde{\mu}_r(z).
\end{equation}
Similar objects $\langle{\cal G}^{\mu},\tilde{\mu}_{\dot{r}}\rangle$ and  $\langle{\cal G},\tilde{\mu}_{\ddot{r}}\rangle$ are defined using an appropriate basis $\{\tilde{\mu}_{\dot{r}}\}$ and $\{\tilde{\mu}_{\ddot{r}}\}$ for the moduli spaces of $e_{\mu}(z)$ and $e(z)$ respectively. Let $D$, $\dot{D}$ and $\ddot{D}$ be the dimensions of the moduli spaces of the fields $e_a(z)$, $e_{\mu}(z)$ and $e(z)$ respectively (i.e. the dimension of the space of such fields, modulo infinitesimal gauge transformations corresponding to the BRST transformations found in the section above). Then $r=1,2,...,D$ and $\ddot{r}=1,2,...,\ddot{D}$ index bosonic directions and $\dot{r}=1,2,...,\dot{D}$ indexes grassmann directions. The ghost systems $(\rho^a,\eta_a)$, $(\rho^{\mu},\eta_{\mu})$ and $(\rho,\eta)$ are of the conventional $b\bar{\partial}c$ type of weight $3/2$, $2$ and $3$ respectively and, neglecting the moduli associated with punctures, we have that $D=(2d-4)(2g-2)$, $\dot{D}=d(3g-3)$ and $\ddot{D}=5g-5$ where $g$ is the genus of $\Sigma$. (The general variation of $e_a(z,\tau)$ is given by $\Delta e_a=-\bar{\partial}\varepsilon_a+f_a^{bc}(\lambda)\varepsilon_ce_b-Z^{\mu}_a\varepsilon_{\mu}+\delta\tau^r\partial_r e_a$, where $\tau^r$ are moduli (coordinates on the space of $e_a$ modulo infinitesimal gauge transformations). We assume the parameters $\varepsilon_a$ and $\varepsilon_{\mu}$ are independent of the moduli $\tau^r$ and so the shift symmetry $e_a\rightarrow e_a-Z^{\mu}_a\varepsilon_{\mu}$ can be used to remove $d-3$ components of $e_a$ for a given $\tau^r$ but the $e_a$ still vary with $\tau^r$, thus the relevant moduli space after gauge-fixing is still $D$-dimensional.) The fibres of ${\cal E}$ are then $(D+\ddot{D}|\dot{D})$-dimensional. Correlation functions involving the $\langle \rho^a,\tilde{\mu}_{r}\rangle$, $\delta(\langle\rho^{\mu},\tilde{\mu}_{\dot{r}}\rangle)$ and $\langle\rho,\tilde{\mu}_{\ddot{r}}\rangle$ are expected to give top (holomorphic) forms on these fibres. More generally, we expect correlation functions involving the $\mathbf{b}(\vec{v}_m)$ in addition to these ghost insertions to have an interpretation as forms on ${\cal E}$. The bundle ${\cal E}$ appears to have a rich structure and more work needs to be done to clarify the details.

\subsection{Gauge Fixing}

Our starting point is the non-minimal action
$$
S=\int_{\Sigma} {\cal Z}\hspace{-.06cm}\cdot\hspace{-.06cm}\bar{\partial}{\cal Z}+e_aG^a +\delta\Psi+\pi^a{\cal F}_a+\pi^{\mu}{\cal F}_{\mu}+\pi {\cal F},
$$
where BRST variations are generated by $\widehat{Q}$, the full BRST charge, which includes the transformations (\ref{d2}), (\ref{d1}) and (\ref{d3}). We take the gauge-fixing fermion to be
\begin{equation}\label{gff}
\Psi=\int_{\Sigma}\rd^2z\Big(\rho^a {\cal F}_a+\rho^{\mu}{\cal F}_{\mu}-\rho {\cal F}\Big),
\end{equation}
where
$$
{\cal F}_a=e_a- \sum_{r=1}^{n-2}s_a^r(z) \tilde{\mu}_r ,	\qquad	{\cal F}_{\mu}=\e_{\mu}- \sum_{\dot{r}}s_{\mu}^{\dot{r}}(z) \tilde{\mu}_{\dot{r}}, \qquad	{\cal F}=\e- \sum_{\ddot{r}}s^{\ddot{r}}(z) \tilde{\mu}_{\ddot{r}}.
$$
The $s_a^r(z)$ are worldsheet fields, which transform as $\delta s_a^r(z)=m_a^r(z)$, where $\delta m_a^r(z)=0$. We introduce similar fields such that $\delta s_{\mu}^{\dot{r}}=m_{\mu}^{\dot{r}}$ and $\delta s^{\ddot{r}}=m^{\ddot{r}}$. The Lagrange multipliers $\pi^a$, $\pi^{\mu}$ and $\pi$ set ${\cal F}_a=0$, ${\cal F}_{\mu}=0$ and ${\cal F}=0$ respectively. Thus the only contribution to the variation of the gauge-fixing fermion $\Psi$ comes from $\delta\Psi\approx-\langle\rho^a,\delta{\cal F}_a\rangle+\langle\rho^{\mu},\delta{\cal F}_{\mu}\rangle+\langle\rho,\delta {\cal F}\rangle$ where it is understood that $\approx$ denotes equality subject to the equations on motion for the $\pi$'s. Using the variations of the Lagrange multipliers $e_a(z)$, $e_{\mu}(z)$ and $e(z)$ derived in the last section, it is straightforward to show that 
\begin{eqnarray}\label{dPsi}
\delta\Psi\approx\int_{\Sigma}\rd^2z \Bigg(&&\rho^a\bar{\partial} \eta_a-\rho^{\mu}\bar{\partial}\eta_{\mu}+\rho\bar{\partial}\eta +\Big({\cal G}^a-G^a\Big)e_a
+{\cal G}^{\mu}e_{\mu}+{\cal G}e \nonumber\\
&&-\sum_ rm_a^r\tilde{\mu}_r\rho^a+\sum_{\dot{r}}m_{\mu}^{\dot{r}}\tilde{\mu}_{\dot{r}}\rho^{\mu}+\sum_{\ddot{r}}m^{\ddot{r}}\tilde{\mu}_{\ddot{r}}\rho\Bigg).
\end{eqnarray}
Integrating out the $m_a^r$, $m_{\mu}^{\dot{r}}$ and $m^{\ddot{r}}$ results in the insertion of the operators 
\begin{equation}\label{B}
\prod_{r,a} \langle\rho^a,\tilde{\mu}_r\rangle,		\qquad	\prod_{\dot{r},\mu}\delta\Big(\langle\rho^{\mu},\tilde{\mu}_{\dot{r}}\rangle\Big),	\qquad	\prod_{\ddot{r}}\langle\rho,\tilde{\mu}_{\ddot{r}}\rangle,
\end{equation}
respectively into the path integral. Similarly, integrating out the $s_a^r(z)$, $s_{\mu}^{\dot{r}}(z)$ and $s^{\ddot{r}}(z)$ results in the insertion of the operators
\begin{equation}\label{GG}
\prod_{a,r}\delta\Big(\langle{\cal G}^a,\tilde{\mu}_{r}\rangle\Big),	\qquad	\prod_{\mu,\dot{r}}\langle{\cal G}^{\mu},\tilde{\mu}_{\dot{r}}\rangle,	\qquad	\prod_{\ddot{r}}\delta\Big(\langle{\cal G},\tilde{\mu}_{\ddot{r}}\rangle\Big).
\end{equation}
Putting the stress tensor ghost contribution in, the gauge-fixed Lagrangian is then
\begin{equation}\label{gfS}
S= \int_{\Sigma} {\cal Z}\hspace{-.06cm}\cdot\hspace{-.06cm}\bar{\partial}{\cal Z} + b\bar{\partial}c+\rho^a\bar{\partial}\eta_a-\rho^{\mu}\bar{\partial}\eta_{\mu}+\rho\bar{\partial}\eta,
\end{equation}
with the ghost insertions discussed above as well as the usual holomorphic $\mathbf{b}(\vec{v}_m)$ insertions. As noted above, the worldsheet vectors $\vec{v}_m(z_i)=(v_1,...,v_{n-3})$ form a basis for the $n-3$ moduli deformations based at the $i$'th puncture $z_i\rightarrow z_i+v(z_i)$. At genus zero the deformations can be chosen to simply translate the puncture. The $SL(2)$ invariance may be used to fix three punctures so that $\vec{v}_m(z_i)=\delta_{m i}$ for $i=1,2,..,n-3$ and vanishes for $i=n-2,n-1,n$ (the three fixed punctures).

Introducing some local operators ${\cal V}_i(z_i)$ in the cohomology of $\widehat{Q}$, a correlation function of tree-level observables is given by\footnote{Due to its statistics, the object
$$
Y=\prod_{\mu,\dot{r}}\langle{\cal G}^{\mu},\tilde{\mu}_{\dot{r}}\rangle\;\delta\Big(\langle\rho^{\mu},\tilde{\mu}_{\dot{r}}\rangle\Big),
$$
plays a role akin to that of a picture changing operator in the supersymmetric theory.}
\begin{equation}\label{integral}
A_n=\int_{\Gamma_n} \left\langle\;\prod_m\mathbf{b}(\vec{v}_m)\;\prod\mathbf{B}(\tilde{\mu})\;\prod\delta \Big(\mathbf{G}(\tilde{\mu})\Big){\cal V}_1...{\cal V}_n\right\rangle.
\end{equation}
where $\prod\mathbf{B}(\tilde{\mu})$ denotes the product of the ghost insertions (\ref{B}) and $\prod\delta \left(\mathbf{G}(\tilde{\mu})\right)$ denotes the product of the terms in (\ref{GG}). The action used to compute the correlation function under the integral is is (\ref{gfS}). $\Gamma_n\subset {\cal E}$ is a cycle that we discuss briefly towards the end of section 4.5, although we freely admit that, at this stage, we have no concrete method of determining it. It may be possible to to adapt the methods of \cite{Ohmori:2015sha} to evaluate the result of this integral in terms of the stationary points of a suitably defined Morse function. 

\subsection{Open algebras and the Master Action}

We finally turn to the fact that the soft algebra only closes on-shell. Starting with the action
$$
S_{(0)}= \int_{\Sigma} {\cal Z}\hspace{-.06cm}\cdot\hspace{-.06cm}\bar{\partial}{\cal Z}+e_aG^a,
$$
we add a gauge fixing term
$$
S_{(1)}=\delta\Psi.
$$
and possibly non-minimal terms involving the $\pi$-fields as above. If the action of the BRST charge is nilpotent then $\delta^2\Psi=0$ and the combined action $S=S_{(0)}+S_{(1)}$ is BRST-invariant. In our case the gauge-fixed action above is not BRST invariant and the origin of this is the fact that the presence of $e_a(z)$ and its descendants in $\Psi$ means that $\delta^2\Psi$ only vanishes up to terms proportional to $\delta S/\delta\omega_a$. The BV construction \cite{Batalin:1984jr, Batalin:1981jr} tells us how to construct an off-shell BRST-invariant action by expanding the space of fields to include anti-fields. The key idea is that, for each field $\phi^i$, one introduces an anti-field\footnote{The $^*$ notation signifies the object to be an antifield. No notion of complex or Hermitian conjugation is intended. We include ghosts in our definition of `field'.} $\phi^*_i$ whose BRST variation is precisely the equation of motion for the corresponding field
$$
\delta \phi^*_i=\frac{\delta {\cal S}}{\delta \phi^i}.
$$
In this way terms proportional to the equations of motion are rendered trivial in the extended BRST cohomology. We clearly need to extend our definitions of the BRST charge $Q$ and classical action to accommodate the anti-fields.

We sketch the basic idea here but details on the formalism may be found in \cite{Henneaux:1989jq} and \cite{Gomis:1994he} contains a number of worked examples. The main new ingredient is the bracket $(\;,\;)$ defined by
$$
({\cal A},{\cal B})=\sum_i\int_{\Sigma}\left(\frac{\delta^r{\cal A}}{\delta \phi^i(z)}\frac{\delta^l {\cal B}}{\delta \phi^*_i(z)}-\frac{\delta^r {\cal A}}{\delta \phi^*_i(z)}\frac{\delta^l {\cal B}}{\delta \phi^i(z)}\right),
$$
where ${\cal A}$ and ${\cal B}$ are functionals of the fields and anti-fields and the $l$ $(r)$ superscript denotes the functional derivatives acting from the left (right), so that for example that $(\phi^i(z),\phi^*_j(w))=\delta^i_j\delta(z-w)$. The classical Master Action ${\cal S}$ is required to satisfy $({\cal S},{\cal S})=0$ and may be found iteratively as a series
$$
{\cal S}=S_{(0)}+S_{(1)}+S_{(2)}+...,
$$
where the `boundary conditions' $S_{(0)}$ and $S_{(1)}$ are, broadly speaking, as above. The expansion may be thought of as a polynomial expansion in antifields where $S_{(0)}$ is the original action containing fields only and $S_{(1)}$ is linear in antifields. The gauge-fixing fermion, which is a functional of the fields, tells us the relationship between the fields and antifields
\begin{equation}\label{anti}
\phi_i^*=\frac{\delta\Psi}{\delta\phi^i},
\end{equation}
and we may write $S_{(1)}=\sum_i\phi_i^*\delta \phi^i$. Indeed, if the algebra closes off-shell then the master action will be at most linear in the antifields
$$
{\cal S}=S_0+\sum_i\frac{\delta\Psi}{\delta\phi^i}\delta \phi^i=S_0+\delta\Psi,
$$
and we can eliminate the antifields entirely. We then extend the minimal gauge-fixed action by introducing terms $b_i^*\pi^i$ such that $\delta b^i=\pi^i$. The BRST transformations, including (\ref{d2}), (\ref{d1}) and (\ref{d3}), are given by
$$
\delta_{\widehat{Q}} \phi^i=({\cal S},\phi^i)=\frac{\delta {\cal S}}{\delta \phi^*_i}.
$$
More generally, the extended notion of BRST transformation is also given by $\delta \phi^i=({\cal S},\phi^i)$ but now the action may have quadratic or higher dependence on the anti-fields.

The ambitwisitor string includes a field $e_a(z)$ whose algebra only closes on-shell and we anticipate a Master Action that is non-linear in the antifields. To illustrate the point, let us see what happens if we try to repeat the previous construction ignoring the fact that the $e_a$ algebra is open. We find that the action $S=S_0+\delta\Psi$ is not BRST invariant. How can this be? The algebra is only nilpotent on the support if the $\lambda^a(z)$ equation of motion. This is a consequence of the fact that the $f_a^{bc}(\lambda)$ are functions of $\lambda^a(z)$ and not constants (and so ultimately the failure of the Octonions to be associative). This problem can be overcome by including non-linear antifield terms in ${\cal S}$. Following the general formalism laid out in \cite{Henneaux:1989jq} it is not too hard to see that what we need is the additional term
$$
S_{(2)}=e^{*a}\omega^{*b}\Big(2\eta_b\eta_a-\Gamma_{ab}^{\mu}\eta_{\mu}\Big)+2e^{*\mu}\omega^{*a}\Big( \Gamma^{\nu ad}\Gamma_{\mu db}\eta_a\eta_{\nu}-\Gamma_{\mu ab}\lambda^b\eta\Big)-4\omega^{*a}e^*\eta_a\eta.
$$
The modified action is then ${\cal S}=S_{(0)}+S_{(1)}+S_{(2)}$. Note that the presence of $e^{a*}(z)$, $e^{*\mu}(z)$, $e^*(z)$ and $\omega^{*a}(z)$ fields will alter the BRST transformation of the $e_a(z)$, $e_{\mu}(z)$, $e(z)$ and $\omega_a(z)$ fields; however, since our chosen gauge-fixing fermion (\ref{gff}) is independent of $\omega_a(z)$, we see that
 $$
\omega^{*a}=\frac{\delta \Psi}{\delta\omega_a}=0,
$$
and so these modifications, though essential in ensuring the BRST-invariance of the theory, do not affect our previous considerations.

As an illustration, let us ignore the reducibility of the symmetries of the theory and focus on the failure of the BRST charge to be nilpotent on $e_a(z)$ (this amounts to setting the $\eta_{\mu}(z)$ and $\eta(z)$ ghosts to zero). The variation of ${\cal S}$ with respect to $\omega^{*a}(z)$ gives the $\lambda^a$ equation of motion (now incorporated into the cohomology as an exact cycle)
$$
\delta \omega^{*a}=({\cal S},\omega^{*a})=\frac{\delta {\cal S}}{\delta\omega_a}=\bar{\partial}\lambda^a+\xi^{ab}e_b.
$$
The $e_a$ transformation is now
$$
\delta e_a=({\cal S},e_a)=\frac{\delta S}{\delta e^*_a}=-\bar{\partial}\eta_a+f_{a}^{bc}\eta_be_c+2\omega^{*b}\eta_a\eta_b.
$$
A quick calculation shows, taking into account the variation of $\omega^{*a}$,  that $\delta^2e_a$ now vanishes.

\subsection{Computing Observables}

We want to move towards computing observables and in this final section we make some speculative remarks in this direction. An important issue is how to make sense of the somewhat formal path integral expression we have derived in previous sections. We will focus on the bosonic case but we expect the supersymmetric generalisation to be straightforward. We will be particularly interested in thinking about the expression (\ref{integral}) as an integral of a top holomorphic form $\Omega=\left\langle {\cal V}_1...{\cal V}_n\right\rangle$ over a bundle ${\cal E}$ with base ${\cal M}$. The correlation function under the integral should be computed using the Master Action subject to the constraint (\ref{anti}). Starting with the gauge-fixed action 
\begin{equation}\label{X}
S=\int_{\Sigma} {\cal Z}\hspace{-.06cm}\cdot\hspace{-.06cm}\bar{\partial}{\cal Z} +S_{\text{gh}}, 
\end{equation}
where $S_{\text{gh}}=b\bar{\partial}c+\rho^a\bar{\partial}\eta_a-\rho^{\mu}\bar{\partial}\eta_{\mu}+\rho\bar{\partial}\eta$, we consider how this changes as we move around the space ${\cal E}$. It is important to note that $\hat{\delta}$ denotes a change in the moduli of the gauge fields, whereas the discussion in  previous sections has largely focussed on infinitesimal gauge transformations at the same point in moduli space. Under a small change in the worldsheet metric $\hat{\delta}\mu=\delta\tau^m(\partial \mu/\partial\tau^m)$, the Lagrangian changes as $\hat{\delta}{\cal L}=\hat{\delta}\mu\, T$ where $T$ is the full stress tensor (\ref{T}). Similarly, a change in the Lagrange multiplier $e_a$ gives rise to $\hat{\delta}{\cal L}=\hat{\delta}e_a\;{\cal G}^a$, where $\hat{\delta}e_a=\delta\tau^r(\partial e_a/\partial\tau^r)$, and so the general variation of $e_a(z,\tau)$ is given by $\Delta e_a=-\bar{\partial}\varepsilon_a-4\delta_a^{[b}\lambda^{c]}\varepsilon_ce_b-Z^{\mu}_a\varepsilon_{\mu}+\delta\tau^r\partial_r e_a$. Similar expressions may be found for changes in $e_{\mu}$ and $e$ and the corresponding response on the gauge-fixed action may be easily deduced from the expression for $\delta\Psi$ given in (\ref{dPsi}), yielding
$$
\hat{\delta} S=\int_{\Sigma}\hat{\delta}\mu\,T+\hat{\delta}e_a\,{\cal G}^a+\hat{\delta}e_{\mu}\,{\cal G}^{\mu}+\hat{\delta}e\,{\cal G}.
$$
Using the transformations (\ref{G}), an invariant action is given by by adding the term \cite{Witten:2012bh}
$$
S_{\text{ext}}=\int_{\Sigma}\,\hat{\delta}\mu \,b+ \hat{\delta} e_a\,\rho^a+ \hat{\delta} e_{\mu}\,\rho^{\mu}+ \hat{\delta}e\,\rho,
$$
to (\ref{X}). This term produces $b$ and $\rho$ ghost zero mode contributions and is another way of seeing how the $\mathbf{b}(\vec{v})$ and $\mathbf{B}(\tilde{\mu})$ insertions (\ref{B}) in (\ref{integral}) arise. Introducing ${\cal W}:=b\,\mu+\rho^a\, e_a+\rho^{\mu}\,e_{\mu}+\rho\, e$, the extended action may be written in terms of the extended BRST operator $\widehat{Q}$ which generates these transformations
$$
S=\int_{\Sigma} {\cal Z}\hspace{-.06cm}\cdot\hspace{-.06cm}\bar{\partial}{\cal Z}+\{\widehat{Q},{\cal W}\}+S_{\text{gh}}.
$$
Given a set of observables ${\cal V}_i(z_i)$, we expect the correlation function of $n$ such observables given by a path integral to follow the general  approach outlined in \cite{Ohmori:2015sha}. The path integral localises on critical points of the Morse function $\Re({\cal I})$ where
$$
{\cal I}=-\langle\mu,T\rangle-\langle e_a,{\cal G}^a\rangle-\langle e_{\mu},{\cal G}^{\mu}\rangle+\langle e,{\cal G}\rangle,
$$
giving the tree-level correlation function as a sum over such critical points. The role of the antifields would need to be carefully examined. The key outstanding task is to identify the cohomology of the BRST operator and to write down concrete examples of vertex operators so that the above prescription may be investigated fully.

\section{Discussion}

There is a sense in which the twistor variables used here provide a more natural description of the ambitwistor string, one that may make the subtleties relating constructions in different dimensions more explicit. We have seen that, even in the bosonic case, working in the twistorial variables ${\cal Z}(z)$ instead of the more familiar pair $(X(z),P(z))$ leads to significant increase in the complexity of the theory due to the more involved constraints. Whilst one might argue that this complexity is indicative of a richness that the twistor variables bring to the surface, it seems likely that explicit computations will be less efficient in this formalism. However, it is possible that certain types of problems that are difficult or intractable in the conventional approach may be fruitfully tackled in this language.

In particular, there has been progress recently in constructing ambitwistor strings in non-trivial Neveu-Schwarz backgrounds \cite{Adamo:2017nia,Adamo:2017sze} and it would also be interesting to generalise the construction here to curved backgrounds. An interesting feature of the target space supersymmetric theory is that, in contrast to the conventional superstring, the reducibility of the constraints of the ambitwistor string considered here appear manageable and one might hope to make progress in the computation of scattering amplitudes with Ramond-Ramond backgrounds. The example of $AdS_5\times S^5$ would be particularly interesting. Even though such a calculation would simply be a re-derivation of known supergravity results one might hope that the method, arising as it does from a worldsheet theory, might shed some light on the corresponding problem in superstring theory.

Before more ambitious applications can be investigated there are a number of outstanding issues to be clarified. The most pressing is to determine the spectrum of the physical states of the supersymmetric theory and to construct explicit vertex operators. The classical starting point is the ambitwistor string of \cite{Mason:2013sva} and so our expectation is that the theory described here also describes Type II supergravity, or at least those results accessible from perturbation theory. Related to this is the fact that one would also like a careful definition of the operators used and possible gauge anomalies. The structure of the ghost vacuum also deserves a more thorough analysis. It would also be interesting to see what the analogue of the scattering equations are in this formalism. We expect the role played by the scattering equations in the conventional formalism will be filled by the constraints $G^a(z)$ but, without concrete expressions for vertex operators in this language, it is difficult to make a concrete proposal. We leave these and other questions to future work.

\begin{center}
\textbf{Acknowledgements}
\end{center}

The authors would like to thank Nathan Berkovits, Lionel Mason and Paul Townsend for helpful discussions.  This work has been partially supported by STFC consolidated grant ST/P000681/1.

\appendix

\section{Gamma Matrices and Octonions}

To write the ten-dimensional constraint in an irreducible form it is necessary to break manifest Lorenz invariance \cite{Cederwall:1992bi}. It is useful to introduce a basis for the octonions $\{e_i\}$, where $1=1,...,7$. In addition we include $e_8=1$ to give and a notion of division algebra conjugation $\bar{e}_i=-e_i$ for $i=1,2,...,7$ and $\bar{e}_8=e_8$. This explicitly breaks Lorenz invariance and the $\Gamma$ matrices may be written as
$$
\Gamma^+_{ab}=\left(\begin{array}{cc} \sqrt{2}\delta^A_B & 0 \\ 0 & 0\end{array}\right),	\qquad	\Gamma^-_{IJ}=\left(\begin{array}{cc} 0 & 0 \\ 0 & \sqrt{2}\delta_A^B\end{array}\right),	\qquad	\Gamma^i_{ab}=\left(\begin{array}{cc} 0 & \sigma^{iAB} \\ -\sigma^i_{AB} & 0\end{array}\right).
$$
where $\sigma^i_{AB}$ ae real antisymmetric Pauli matrices of $SO(7)$ and we define $\sigma^8_{AB}=\delta_{AB}$. They satsify $\{\sigma^i,\sigma^j\}=-2\delta^{ij}$.

It is useful to introduce octonionic vielbeins to map us from the $SL(16;\R)$ representation above to the $SL(2;\mathbb{O})$. In terms of the octonions $e_A$, we introduce $E^a_I=\left( e_A , e_A\right)$, so that the sixteen components of $\lambda^a$ may be written as a two component $SL(2;\mathbb{O})$ spinor $\lambda_I=E_{Ia}\lambda^a$, which we write as
$\lambda_I=\left(	\lambda_1 , \lambda_2\right)$, where we set $\lambda^a=(\lambda_1^A,\lambda_2^A)$, where $A=1,2,...8$ and so
$$
\lambda_1=\sum_{A=1}^8\lambda_1^Ae_A,	\qquad	\lambda_2=\sum_{A=1}^8\lambda_2^Ae_A,
$$
Similarly, the gamma matrices may be written as $SL(2;\mathbb{O})$ bispinors $\Gamma^{\mu}_{IJ}=\frac{1}{8}\Gamma^{\mu}_{ab}E^a_I\bar{E^b_J}$. Explicitly, we have \cite{Cederwall:1992bi}
$$
\Gamma^+_{IJ}=\left(\begin{array}{cc} \sqrt{2} & 0 \\ 0 & 0\end{array}\right),	\qquad	\Gamma^-_{IJ}=\left(\begin{array}{cc} 0 & 0 \\ 0 & \sqrt{2}\end{array}\right),	\qquad	\Gamma^i_{IJ}=\left(\begin{array}{cc} 0 & e_i \\ \bar{e}^i & 0\end{array}\right),	\qquad	\Gamma^8_{IJ}=\left(\begin{array}{cc} 0 & 1 \\ 1 & 0\end{array}\right).
$$
These satisfy the Clifford algebra relation and care must be taken to keep track of the order of operations as the Octonions are not associative. It will helpful to combine $\Gamma^A=(\Gamma^i,\Gamma^8)$, where $A=1,2,..8$.

We are interested in transformations of the $\lambda's$ that preserve the momentum $P_{\mu}$. We can write the momentum in $SL(2;\mathbb{O})$ notation as $P_{IJ}=\Gamma_{IJ}^{\mu}P_{\mu}$ where
$$
P_{IJ}=\left(\begin{array}{cc} \sqrt{2}P^+ & P^Ae_A \\  P^A\bar{e}_A & \sqrt{2}P^-  \end{array}\right)
$$
where $I=1,2$. We can write $P_{IJ}=\lambda_I\bar{\lambda}_J$
$$
P_{IJ}=\left(\begin{array}{cc} |\lambda_2|^2 & \lambda_1\bar{\lambda}_2 \\ \lambda_2\bar{\lambda}_1 & |\lambda_1|^2 \end{array}\right)
$$
It is clear that $P^2=\det(P_{IJ})=0$.

\section{Critical dimension}

All fields in the flat space theory are free. In particular they have kinetic terms of the form of a $\beta\gamma$ system. As such we can import the standard results that field of weight $\ell$ and Grassmann parity $\epsilon$ ($\epsilon=+1$ for bosons and $-1$ for fermions) contributes
$$
N_{\ell}c_{\ell}=\frac{1}{24}\epsilon [3(2\ell-1)^2-1],
$$
to the central charge, where $N_{\ell}$ is the number of such fields.

\vspace{.5cm}

\begin{tabular}{ |p{2cm}||p{2cm}|p{2cm}|p{2.8cm}|p{2cm}|  }
 \hline
 \multicolumn{5}{|c|}{Central Charge Contributions} \\
 \hline
 Field  & Weight $\ell$ & Statistics $\epsilon$ &  Multiplicity $N_{\ell}$ & $24\times c_{\ell}$ \\
 \hline
 $(\omega_a,\lambda^a)$   & $\frac{1}{2}$ & $+1$ &   $2d-4$ & $-1$\\
 $(b,c)$ &   $2$  & $-1$ & $1$  & $-26$ \\
 $(\rho^a,\eta_a)$ & $\frac{3}{2}$ & $-1$ & $2d-4$ &  $-11$\\
 $(\rho^{\mu},\eta_{\mu})$ & $2$ & $+1$ & $d$ &  $+26$\\
 $(\rho,\eta)$ & $3$ & $-1$ & $1$ &  $-74$\\
 \hline
\end{tabular}

\vspace{.5cm}

The total central charge is then
$$
c=\sum_{\ell}N_{\ell}c_{\ell}=\frac{1}{24}\Big(-(2d-4)-26-11(2d-4)+26 d-74\Big)=\frac{2(d-26)}{24}.
$$
This vanishes in the critical dimension $d=26$.

\section{Comments on Supersymmetry}

We do not expect the supersymmetric extension to the bosonic BRST charge to present any novel difficulty. An interesting feature is that the $\kappa_a$ variable in the target space supersymmetric formulation only appears in the combination $\varepsilon=\lambda^a\kappa_a$ and the structure of the supersymmetry transformations in RNS and Green-Schwarz constructions is identical. It should be noted that the target space and the constraints that appear in each version are  very different. Nonetheless this suggests that the twistor formulation may provide a better formalism in which make contact between the two versions of the superstring.

\subsection{Target Space Supersymmetry}

Our starting point is the Green-Schwarz action
$$
S=\int_{\Sigma}P_{\mu}(\bar{\partial}X^{\mu}-\Gamma^{\mu}_{ab}\theta_r^a\bar{\partial}\theta_r^b)+\frac{1}{2}hP^2,	\qquad	r=1,2.
$$
where $\theta^a_r$ is a grassman scalar on the worldsheet. The theory is invariant under $\kappa$ symmetry.
$$
\delta \theta^a=P^{\mu}\Gamma_{\mu}^{ab}\kappa_b,	\qquad	\delta P^{\mu}=0,	\qquad	\delta X^{\mu}=-i\theta^a\Gamma^{\mu}_{ab}(\delta\theta^b),	\qquad	\delta h=-4i(\bar{\partial}\theta^a)\kappa_a,
$$
where $\kappa^a$ is a local fermionic field.

We now turn to the ambitwistor description of the theory. The incidence relation is modified to
$$
\mu_a=X_{\mu}\Gamma^{\mu}_{ab}\lambda^b-i\psi_{\mu}\Gamma^{\mu}_{ab}\theta^b,	\qquad	\psi^{\mu}_r=\Gamma^{\mu}_{ab}\theta^a_r\lambda^b,
$$
with the latter solving the constraints $P_{\mu}\psi^{\mu}_r=0$. The bosonic constraint becomes
$$
{\cal K}^a=G^a+2i\psi^{\mu}\psi_{\nu}\Gamma_{\mu}^{ac}\Gamma^{\nu}_{cb}\lambda^b
$$
There is an additional pair of (irreducible) constraints
$$
S_r=\lambda^a\Gamma_{\mu ab}\lambda^b\psi_r^{\mu}.
$$
We introduce Lagrange multipliers $h^r$ (of weight $(-1/2,1)$) to impose these constraints and so the action is
$$
S=\frac{1}{2}\int_{\Sigma}\mu_a\bar{\partial}\lambda^a-\lambda^a\bar{\partial}\mu_a+i\tilde{\psi}_{\mu}\bar{\partial}\psi^{\nu}+\mu T+h_a{\cal K}^a+h^rS_r
$$
Kappa symmetry preserves this action. The constraints $S_r$ transform trivially as $\delta S_r=2(\lambda^a\kappa_a^r)P^2=0$; however, the bosonic constraint has non-trivial transformation
$\delta{\cal K}^a=8i(\lambda^b\kappa_b^r)\lambda^aS_r$, indicating that the Lagrange multiplier $h_r$ has a non-trivial variation. The fields transform as
$$
\delta\lambda^a=0,	\qquad	\delta \psi^{\mu}_r=2(\lambda^a\kappa^r_a)P^{\mu},	\qquad	\delta \mu_a=8i(\lambda^b\kappa^r_b)\Gamma^{\mu}_{ac}\psi_r^{\mu}\lambda^c, 
$$
$$
\delta h_a=0,	\qquad	\delta h^r=8i(\lambda^b\kappa_b^r)h_a\lambda^a.
$$
We see that the parameter only appears in the combination $\lambda^a\kappa_a^r$.

\subsection{Worldsheet Supersymmetry}

We now consider the RNS formulation of the theory and include additional Grassmann fields. The incidence relation is unchanged from the bosonic case. but we have the addition of a supercurrent and worldsheet gravitini $\chi_r$
$$
S=\int_{\Sigma}P_{\mu}\bar{\partial}X^{\mu}+i\eta_{\mu\nu}\psi^{\mu}\bar{\partial}\psi^{\nu}+\mu T+P_{\mu}\psi^{\mu}\chi+\frac{1}{2}hP^2.
$$
The worldsheet gravitini $\chi_r$ are Lagrange multipliers for the constraints
$$
S_r=P_{\mu}\psi^{\mu}_r=0.
$$
The theory is invariant under worldsheet supersymmetry
$$
\delta X^{\mu}=\epsilon^r\psi^{\mu}_r,	\qquad	\delta P_{\mu}=0,	\qquad	\delta\psi^{\mu}_r=-2\epsilon_rP^{\mu},
$$
$$
\delta h=4\epsilon^r\chi_r,	\qquad	\delta \chi_r=2\bar{\partial}\epsilon.
$$
We introduce $\lambda^a$ in the standard way and imposing the incidence relations
$$
\omega_a=X_{\mu}\Gamma^{\mu}_{ab}\lambda^b.
$$
The constraint $G^a$ is the same as in the bosonic theory. The spinorial action is
$$
S=\frac{1}{2}\int_{\Sigma}\omega_a\bar{\partial}\lambda^a-\lambda^a\bar{\partial}\omega_a+i\eta_{\mu\nu}\psi^{\mu}\bar{\partial}\psi^{\nu}+\mu T+e_aG^a+S_r\chi^r
$$
The action is invariant under local supersymmetries
$$
\delta\lambda^a=0,	\qquad \delta \psi_r^{\mu}=-2\epsilon (\lambda^a\Gamma^{\mu}_{ab}\lambda^b),	\qquad	\delta\omega_a=\epsilon \psi_r^{\mu}\Gamma_{\mu ab}\lambda^b
$$
$$
\delta e_a=0,	\qquad	\delta \chi_r=2\bar{\partial}\epsilon.
$$

\subsection{The central charge}

Intriguingly, both the Green-Schwarz and the RNS strings have the same field content. The key difference are the constraints. We conjecture that the imposition of the $(G^a,S)$ constraints, followed by a GSO projection is equivalent to the imposition of the $({\cal K}^a,S)$ constraints.

We note that the worldsheet spinors $\psi^{\mu}$ are taken to be complex valued (i.e. two copies of the real spinor). As such there are two currents; $S=\lambda^a\Gamma_{\mu ab}\lambda^b\psi^{\mu}$ and $\widetilde{S}=\lambda^a\Gamma_{\mu ab}\lambda^b\bar{\psi}^{\mu}$ and associated $(\beta,\gamma)$ and $(\tilde{\beta},\tilde{\gamma})$ ghost systems.

\vspace{.5cm}

\begin{tabular}{ |p{2cm}||p{2cm}|p{2cm}|p{2.8cm}|p{2cm}|  }
 \hline
 \multicolumn{5}{|c|}{Central Charge Contributions} \\
 \hline
 Field  & Weight $\ell$ & Statistics $\epsilon$ &  Multiplicity $N_{\ell}$ & $24\times c_{\ell}$ \\
 \hline
 $(\omega_a,\lambda^a)$   & $\frac{1}{2}$ & $+1$ &   $2d-4$ & $-1$\\
 $(\bar{\psi}_{\mu},\psi^{\mu})$   & $\frac{1}{2}$ & $-1$ &   $d$ & $+1$\\
 $(b,c)$ &   $2$  & $-1$ & $1$  & $-26$ \\
 $(\rho^a,\eta_a)$ & $\frac{3}{2}$ & $-1$ & $2d-4$ &  $-11$\\
 $(\rho^{\mu},\eta_{\mu})$ & $2$ & $+1$ & $d$ &  $+26$\\
 $(\rho,\eta)$ & $3$ & $-1$ & $1$ &  $-74$\\
 $(\beta,\gamma)$ & $\frac{3}{2}$ & $+1$ & $1$ &  $+11$\\
$(\tilde{\beta},\tilde{\gamma})$ & $\frac{3}{2}$ & $+1$ & $1$ &  $+11$\\
 \hline
\end{tabular}

\vspace{.5cm}

The total central charge is then
$$
c=\frac{1}{24}\Big(-(2d-4)+d-26-11(2d-4)+26 d-74+11+11\Big)=\frac{3(d-10)}{24}.
$$
This vanishes in the critical dimension $d=10$.



\end{document}